\documentclass[longauth]{aa}  

\usepackage{graphicx,xcolor,euclid}
\usepackage{hyperref}
\usepackage{amsmath,amssymb,amsfonts}
\usepackage{txfonts}
\usepackage{tabularx}
\usepackage{multirow} 
\usepackage{changepage}

\usepackage[switch, modulo]{lineno}
\usepackage{ulem}

\let\Gamma\varGamma
\let\Delta\varDelta
\let\Theta\varTheta
\let\Xi\varXi
\let\Pi\varPi
\let\Sigma\varSigma
\let\Upsilon\varUpsilon
\let\Phi\varPhi
\let\Psi\varPsi
\let\Omega\varOmega

\newcommand{\cosmosis}{\texttt{CosmoSIS}}

\newcommand{\emcee}{\texttt{emcee}}

\newcommand{\camb}{\texttt{CAMB}}

\newcommand{\de}{\mathrm{d}}

\newcommand{\om}{\Omega_{{\rm m},0}}
\newcommand{\ob}{\Omega_{{\rm b},0}}

\newcommand{\orcid}[1]{} 
\definecolor{mordantred19}{rgb}{0.68, 0.05, 0.0}

\usepackage{comment}

\usepackage[nameinlink,noabbrev]{cleveref}
\Crefname{equation}{Eq.}{Eqs.}
\Crefname{section}{Sect.}{Sects.}
\Crefname{figure}{Fig.}{Figs.}
\crefname{equation}{Equation}{Equations}
\crefname{section}{Section}{Sections}
\crefname{figure}{Figure}{Figures}

\defcitealias{ISTF}{EP:VII}
\defcitealias{Pocino2021}{EP:XII}

\begin{document}

   \title{\Euclid preparation. XXXIV. The effect of linear redshift-space distortions in photometric galaxy clustering and its cross-correlation with cosmic shear} 
   \titlerunning{\Euclid: RSDs in GCph and the 3$\times$2pt for {\it Euclid}}
   



\author{Euclid Collaboration: K.~Tanidis$^{1,2}$\thanks{\email{konstantinos.tanidis@physics.ox.ac.uk}}, V.~F.~Cardone$^{3,4}$, M.~Martinelli\orcid{0000-0002-6943-7732}$^{3,4}$, I.~Tutusaus\orcid{0000-0002-3199-0399}$^{5,6}$, S.~Camera\orcid{0000-0003-3399-3574}$^{7,8,9}$, N.~Aghanim$^{10}$, A.~Amara$^{11}$, S.~Andreon\orcid{0000-0002-2041-8784}$^{12}$, N.~Auricchio\orcid{0000-0003-4444-8651}$^{13}$, M.~Baldi\orcid{0000-0003-4145-1943}$^{14,13,15}$
% ...
}

\institute{$^{1}$ CEICO, Institute of Physics of the Czech Academy of Sciences, Na Slovance 2, Praha 8, Czech Republic\\
$^{2}$ Department of Physics, Oxford University, Keble Road, Oxford OX1 3RH, UK\\
$^{3}$ INAF-Osservatorio Astronomico di Roma, Via Frascati 33, 00078 Monteporzio Catone, Italy\\
$^{4}$ INFN-Sezione di Roma, Piazzale Aldo Moro, 2 - c/o Dipartimento di Fisica, Edificio G. Marconi, 00185 Roma, Italy\\
$^{5}$ Institut de Recherche en Astrophysique et Plan\'etologie (IRAP), Universit\'e de Toulouse, CNRS, UPS, CNES, 14 Av. Edouard Belin, 31400 Toulouse, France\\
$^{6}$ Universit\'e de Gen\`eve, D\'epartement de Physique Th\'eorique and Centre for Astroparticle Physics, 24 quai Ernest-Ansermet, CH-1211 Gen\`eve 4, Switzerland\\
$^{7}$ Dipartimento di Fisica, Universit\'a degli Studi di Torino, Via P. Giuria 1, 10125 Torino, Italy\\
$^{8}$ INFN-Sezione di Torino, Via P. Giuria 1, 10125 Torino, Italy\\
$^{9}$ INAF-Osservatorio Astrofisico di Torino, Via Osservatorio 20, 10025 Pino Torinese (TO), Italy\\
$^{10}$ Universit\'e Paris-Saclay, CNRS, Institut d'astrophysique spatiale, 91405, Orsay, France\\
$^{11}$ Institute of Cosmology and Gravitation, University of Portsmouth, Portsmouth PO1 3FX, UK\\
$^{12}$ INAF-Osservatorio Astronomico di Brera, Via Brera 28, 20122 Milano, Italy\\
$^{13}$ INAF-Osservatorio di Astrofisica e Scienza dello Spazio di Bologna, Via Piero Gobetti 93/3, 40129 Bologna, Italy\\
$^{14}$ Dipartimento di Fisica e Astronomia, Universit\'a di Bologna, Via Gobetti 93/2, 40129 Bologna, Italy\\
$^{15}$ INFN-Sezione di Bologna, Viale Berti Pichat 6/2, 40127 Bologna, Italy\\
$^{16}$ Dipartimento di Fisica, Universit\'a di Genova, Via Dodecaneso 33, 16146, Genova, Italy\\
$^{17}$ INFN-Sezione di Genova, Via Dodecaneso 33, 16146, Genova, Italy\\
$^{18}$ Department of Physics "E. Pancini", University Federico II, Via Cinthia 6, 80126, Napoli, Italy\\
$^{19}$ INAF-Osservatorio Astronomico di Capodimonte, Via Moiariello 16, 80131 Napoli, Italy\\
$^{20}$ Instituto de Astrof\'isica e Ci\^encias do Espa\c{c}o, Universidade do Porto, CAUP, Rua das Estrelas, PT4150-762 Porto, Portugal\\
$^{21}$ INAF-IASF Milano, Via Alfonso Corti 12, 20133 Milano, Italy\\
$^{22}$ Institut de F\'{i}sica d'Altes Energies (IFAE), The Barcelona Institute of Science and Technology, Campus UAB, 08193 Bellaterra (Barcelona), Spain\\
$^{23}$ Port d'Informaci\'{o} Cient\'{i}fica, Campus UAB, C. Albareda s/n, 08193 Bellaterra (Barcelona), Spain\\
$^{24}$ Institute for Theoretical Particle Physics and Cosmology (TTK), RWTH Aachen University, 52056 Aachen, Germany\\
$^{25}$ INFN section of Naples, Via Cinthia 6, 80126, Napoli, Italy\\
$^{26}$ Dipartimento di Fisica e Astronomia "Augusto Righi" - Alma Mater Studiorum Universit\'a di Bologna, Viale Berti Pichat 6/2, 40127 Bologna, Italy\\
$^{27}$ Centre National d'Etudes Spatiales -- Centre spatial de Toulouse, 18 avenue Edouard Belin, 31401 Toulouse Cedex 9, France\\
$^{28}$ Institut national de physique nucl\'eaire et de physique des particules, 3 rue Michel-Ange, 75794 Paris C\'edex 16, France\\
$^{29}$ Institute for Astronomy, University of Edinburgh, Royal Observatory, Blackford Hill, Edinburgh EH9 3HJ, UK\\
$^{30}$ European Space Agency/ESRIN, Largo Galileo Galilei 1, 00044 Frascati, Roma, Italy\\
$^{31}$ ESAC/ESA, Camino Bajo del Castillo, s/n., Urb. Villafranca del Castillo, 28692 Villanueva de la Ca\~nada, Madrid, Spain\\
$^{32}$ University of Lyon, Univ Claude Bernard Lyon 1, CNRS/IN2P3, IP2I Lyon, UMR 5822, 69622 Villeurbanne, France\\
$^{33}$ Institute of Physics, Laboratory of Astrophysics, Ecole Polytechnique F\'ed\'erale de Lausanne (EPFL), Observatoire de Sauverny, 1290 Versoix, Switzerland\\
$^{34}$ UCB Lyon 1, CNRS/IN2P3, IUF, IP2I Lyon, 4 rue Enrico Fermi, 69622 Villeurbanne, France\\
$^{35}$ Departamento de F\'isica, Faculdade de Ci\^encias, Universidade de Lisboa, Edif\'icio C8, Campo Grande, PT1749-016 Lisboa, Portugal\\
$^{36}$ Instituto de Astrof\'isica e Ci\^encias do Espa\c{c}o, Faculdade de Ci\^encias, Universidade de Lisboa, Campo Grande, 1749-016 Lisboa, Portugal\\
$^{37}$ Department of Astronomy, University of Geneva, ch. d'Ecogia 16, 1290 Versoix, Switzerland\\
$^{38}$ INFN-Padova, Via Marzolo 8, 35131 Padova, Italy\\
$^{39}$ INAF-Istituto di Astrofisica e Planetologia Spaziali, via del Fosso del Cavaliere, 100, 00100 Roma, Italy\\
$^{40}$ Universit\'e Paris-Saclay, Universit\'e Paris Cit\'e, CEA, CNRS, AIM, 91191, Gif-sur-Yvette, France\\
$^{41}$ Institute of Space Sciences (ICE, CSIC), Campus UAB, Carrer de Can Magrans, s/n, 08193 Barcelona, Spain\\
$^{42}$ Institut d'Estudis Espacials de Catalunya (IEEC), Carrer Gran Capit\'a 2-4, 08034 Barcelona, Spain\\
$^{43}$ INAF-Osservatorio Astronomico di Trieste, Via G. B. Tiepolo 11, 34143 Trieste, Italy\\
$^{44}$ Aix-Marseille Universit\'e, CNRS/IN2P3, CPPM, Marseille, France\\
$^{45}$ INAF-Osservatorio Astronomico di Padova, Via dell'Osservatorio 5, 35122 Padova, Italy\\
$^{46}$ Max Planck Institute for Extraterrestrial Physics, Giessenbachstr. 1, 85748 Garching, Germany\\
$^{47}$ University Observatory, Faculty of Physics, Ludwig-Maximilians-Universit{\"a}t, Scheinerstr. 1, 81679 Munich, Germany\\
$^{48}$ Dipartimento di Fisica "Aldo Pontremoli", Universit\'a degli Studi di Milano, Via Celoria 16, 20133 Milano, Italy\\
$^{49}$ INFN-Sezione di Milano, Via Celoria 16, 20133 Milano, Italy\\
$^{50}$ Institute of Theoretical Astrophysics, University of Oslo, P.O. Box 1029 Blindern, 0315 Oslo, Norway\\
$^{51}$ Jet Propulsion Laboratory, California Institute of Technology, 4800 Oak Grove Drive, Pasadena, CA, 91109, USA\\
$^{52}$ Department of Physics, Lancaster University, Lancaster, LA1 4YB, UK\\
$^{53}$ Technical University of Denmark, Elektrovej 327, 2800 Kgs. Lyngby, Denmark\\
$^{54}$ Cosmic Dawn Center (DAWN), Denmark\\
$^{55}$ Max-Planck-Institut f\"ur Astronomie, K\"onigstuhl 17, 69117 Heidelberg, Germany\\
$^{56}$ Department of Physics and Astronomy, University College London, Gower Street, London WC1E 6BT, UK\\
$^{57}$ Department of Physics and Helsinki Institute of Physics, Gustaf H\"allstr\"omin katu 2, 00014 University of Helsinki, Finland\\
$^{58}$ Department of Physics, P.O. Box 64, 00014 University of Helsinki, Finland\\
$^{59}$ Helsinki Institute of Physics, Gustaf H{\"a}llstr{\"o}min katu 2, University of Helsinki, Helsinki, Finland\\
$^{60}$ NOVA optical infrared instrumentation group at ASTRON, Oude Hoogeveensedijk 4, 7991PD, Dwingeloo, The Netherlands\\
$^{61}$ Universit\"at Bonn, Argelander-Institut f\"ur Astronomie, Auf dem H\"ugel 71, 53121 Bonn, Germany\\
$^{62}$ Aix-Marseille Universit\'e, CNRS, CNES, LAM, Marseille, France\\
$^{63}$ Dipartimento di Fisica e Astronomia "Augusto Righi" - Alma Mater Studiorum Universit\'a di Bologna, via Piero Gobetti 93/2, 40129 Bologna, Italy\\
$^{64}$ Department of Physics, Institute for Computational Cosmology, Durham University, South Road, DH1 3LE, UK\\
$^{65}$ Universit\'e C\^{o}te d'Azur, Observatoire de la C\^{o}te d'Azur, CNRS, Laboratoire Lagrange, Bd de l'Observatoire, CS 34229, 06304 Nice cedex 4, France\\
$^{66}$ Universit\'e Paris Cit\'e, CNRS, Astroparticule et Cosmologie, 75013 Paris, France\\
$^{67}$ European Space Agency/ESTEC, Keplerlaan 1, 2201 AZ Noordwijk, The Netherlands\\
$^{68}$ Department of Physics and Astronomy, University of Aarhus, Ny Munkegade 120, DK-8000 Aarhus C, Denmark\\
$^{69}$ Centre for Astrophysics, University of Waterloo, Waterloo, Ontario N2L 3G1, Canada\\
$^{70}$ Department of Physics and Astronomy, University of Waterloo, Waterloo, Ontario N2L 3G1, Canada\\
$^{71}$ Perimeter Institute for Theoretical Physics, Waterloo, Ontario N2L 2Y5, Canada\\
$^{72}$ Universit\'e Paris-Saclay, Universit\'e Paris Cit\'e, CEA, CNRS, Astrophysique, Instrumentation et Mod\'elisation Paris-Saclay, 91191 Gif-sur-Yvette, France\\
$^{73}$ Space Science Data Center, Italian Space Agency, via del Politecnico snc, 00133 Roma, Italy\\
$^{74}$ CEA Saclay, DFR/IRFU, Service d'Astrophysique, Bat. 709, 91191 Gif-sur-Yvette, France\\
$^{75}$ Institute of Space Science, Str. Atomistilor, nr. 409 M\u{a}gurele, Ilfov, 077125, Romania\\
$^{76}$ Dipartimento di Fisica e Astronomia "G. Galilei", Universit\'a di Padova, Via Marzolo 8, 35131 Padova, Italy\\
$^{77}$ Universit\"ats-Sternwarte M\"unchen, Fakult\"at f\"ur Physik, Ludwig-Maximilians-Universit\"at M\"unchen, Scheinerstrasse 1, 81679 M\"unchen, Germany\\
$^{78}$ Departamento de F\'isica, FCFM, Universidad de Chile, Blanco Encalada 2008, Santiago, Chile\\
$^{79}$ Satlantis, University Science Park, Sede Bld 48940, Leioa-Bilbao, Spain\\
$^{80}$ Centro de Investigaciones Energ\'eticas, Medioambientales y Tecnol\'ogicas (CIEMAT), Avenida Complutense 40, 28040 Madrid, Spain\\
$^{81}$ Instituto de Astrof\'isica e Ci\^encias do Espa\c{c}o, Faculdade de Ci\^encias, Universidade de Lisboa, Tapada da Ajuda, 1349-018 Lisboa, Portugal\\
$^{82}$ Universidad Polit\'ecnica de Cartagena, Departamento de Electr\'onica y Tecnolog\'ia de Computadoras,  Plaza del Hospital 1, 30202 Cartagena, Spain\\
$^{83}$ Kapteyn Astronomical Institute, University of Groningen, PO Box 800, 9700 AV Groningen, The Netherlands\\
$^{84}$ INFN-Bologna, Via Irnerio 46, 40126 Bologna, Italy\\
$^{85}$ Infrared Processing and Analysis Center, California Institute of Technology, Pasadena, CA 91125, USA\\
$^{86}$ IFPU, Institute for Fundamental Physics of the Universe, via Beirut 2, 34151 Trieste, Italy\\
$^{87}$ Instituto de Astrof\'isica de Canarias, Calle V\'ia L\'actea s/n, 38204, San Crist\'obal de La Laguna, Tenerife, Spain\\
$^{88}$ University of Applied Sciences and Arts of Northwestern Switzerland, School of Engineering, 5210 Windisch, Switzerland\\
$^{89}$ Institut d'Astrophysique de Paris, 98bis Boulevard Arago, 75014, Paris, France\\
$^{90}$ Junia, EPA department, 41 Bd Vauban, 59800 Lille, France\\
$^{91}$ Instituto de F\'isica Te\'orica UAM-CSIC, Campus de Cantoblanco, 28049 Madrid, Spain\\
$^{92}$ CERCA/ISO, Department of Physics, Case Western Reserve University, 10900 Euclid Avenue, Cleveland, OH 44106, USA\\
$^{93}$ Laboratoire de Physique de l'\'Ecole Normale Sup\'erieure, ENS, Universit\'e PSL, CNRS, Sorbonne Universit\'e, 75005 Paris, France\\
$^{94}$ Observatoire de Paris, Universit\'e PSL, Sorbonne Universit\'e, LERMA, 750 Paris, France\\
$^{95}$ Astrophysics Group, Blackett Laboratory, Imperial College London, London SW7 2AZ, UK\\
$^{96}$ Scuola Normale Superiore, Piazza dei Cavalieri 7, 56126 Pisa, Italy\\
$^{97}$ SISSA, International School for Advanced Studies, Via Bonomea 265, 34136 Trieste TS, Italy\\
$^{98}$ INFN, Sezione di Trieste, Via Valerio 2, 34127 Trieste TS, Italy\\
$^{99}$ Departamento de Astrof\'isica, Universidad de La Laguna, 38206, La Laguna, Tenerife, Spain\\
$^{100}$ Dipartimento di Fisica e Scienze della Terra, Universit\'a degli Studi di Ferrara, Via Giuseppe Saragat 1, 44122 Ferrara, Italy\\
$^{101}$ Istituto Nazionale di Fisica Nucleare, Sezione di Ferrara, Via Giuseppe Saragat 1, 44122 Ferrara, Italy\\
$^{102}$ Institut de Physique Th\'eorique, CEA, CNRS, Universit\'e Paris-Saclay 91191 Gif-sur-Yvette Cedex, France\\
$^{103}$ Institut d'Astrophysique de Paris, UMR 7095, CNRS, and Sorbonne Universit\'e, 98 bis boulevard Arago, 75014 Paris, France\\
$^{104}$ Dipartimento di Fisica - Sezione di Astronomia, Universit\'a di Trieste, Via Tiepolo 11, 34131 Trieste, Italy\\
$^{105}$ NASA Ames Research Center, Moffett Field, CA 94035, USA\\
$^{106}$ Kavli Institute for Particle Astrophysics \& Cosmology (KIPAC), Stanford University, Stanford, CA 94305, USA\\
$^{107}$ INAF, Istituto di Radioastronomia, Via Piero Gobetti 101, 40129 Bologna, Italy\\
$^{108}$ Institute Lorentz, Leiden University, PO Box 9506, Leiden 2300 RA, The Netherlands\\
$^{109}$ Institute for Astronomy, University of Hawaii, 2680 Woodlawn Drive, Honolulu, HI 96822, USA\\
$^{110}$ Department of Physics \& Astronomy, University of California Irvine, Irvine CA 92697, USA\\
$^{111}$ Departamento F\'isica Aplicada, Universidad Polit\'ecnica de Cartagena, Campus Muralla del Mar, 30202 Cartagena, Murcia, Spain\\
$^{112}$ Department of Astronomy \& Physics and Institute for Computational Astrophysics, Saint Mary's University, 923 Robie Street, Halifax, Nova Scotia, B3H 3C3, Canada\\
$^{113}$ Dipartimento di Fisica, Universit\'a degli studi di Genova, and INFN-Sezione di Genova, via Dodecaneso 33, 16146, Genova, Italy\\
$^{114}$ Department of Computer Science, Aalto University, PO Box 15400, Espoo, FI-00 076, Finland\\
$^{115}$ Ruhr University Bochum, Faculty of Physics and Astronomy, Astronomical Institute (AIRUB), German Centre for Cosmological Lensing (GCCL), 44780 Bochum, Germany\\
$^{116}$ Universit\'e Paris-Saclay, CNRS/IN2P3, IJCLab, 91405 Orsay, France\\
$^{117}$ Department of Physics and Astronomy, Vesilinnantie 5, 20014 University of Turku, Finland\\
$^{118}$ Serco for European Space Agency (ESA), Camino bajo del Castillo, s/n, Urbanizacion Villafranca del Castillo, Villanueva de la Ca\~nada, 28692 Madrid, Spain\\
$^{119}$ AIM, CEA, CNRS, Universit\'{e} Paris-Saclay, Universit\'{e} de Paris, 91191 Gif-sur-Yvette, France\\
$^{120}$ Oskar Klein Centre for Cosmoparticle Physics, Department of Physics, Stockholm University, Stockholm, SE-106 91, Sweden\\
$^{121}$ Univ. Grenoble Alpes, CNRS, Grenoble INP, LPSC-IN2P3, 53, Avenue des Martyrs, 38000, Grenoble, France\\
$^{122}$ Centre de Calcul de l'IN2P3/CNRS, 21 avenue Pierre de Coubertin 69627 Villeurbanne Cedex, France\\
$^{123}$ Dipartimento di Fisica, Sapienza Universit\`a di Roma, Piazzale Aldo Moro 2, 00185 Roma, Italy\\
$^{124}$ Centro de Astrof\'{\i}sica da Universidade do Porto, Rua das Estrelas, 4150-762 Porto, Portugal\\
$^{125}$ Dipartimento di Fisica, Universit\`a di Roma Tor Vergata, Via della Ricerca Scientifica 1, Roma, Italy\\
$^{126}$ INFN, Sezione di Roma 2, Via della Ricerca Scientifica 1, Roma, Italy\\
$^{127}$ Department of Mathematics and Physics E. De Giorgi, University of Salento, Via per Arnesano, CP-I93, 73100, Lecce, Italy\\
$^{128}$ INAF-Sezione di Lecce, c/o Dipartimento Matematica e Fisica, Via per Arnesano, 73100, Lecce, Italy\\
$^{129}$ INFN, Sezione di Lecce, Via per Arnesano, CP-193, 73100, Lecce, Italy\\
$^{130}$ Institute for Computational Science, University of Zurich, Winterthurerstrasse 190, 8057 Zurich, Switzerland\\
$^{131}$ Institut f\"ur Theoretische Physik, University of Heidelberg, Philosophenweg 16, 69120 Heidelberg, Germany\\
$^{132}$ Universit\'e St Joseph; Faculty of Sciences, Beirut, Lebanon\\
$^{133}$ Mullard Space Science Laboratory, University College London, Holmbury St Mary, Dorking, Surrey RH5 6NT, UK\\
$^{134}$ Department of Astrophysical Sciences, Peyton Hall, Princeton University, Princeton, NJ 08544, USA\\
$^{135}$ Niels Bohr Institute, University of Copenhagen, Jagtvej 128, 2200 Copenhagen, Denmark\\
$^{136}$ Cosmic Dawn Center (DAWN)\\
$^{137}$ Universit\"at Innsbruck, Institut f\"ur Astro- und Teilchenphysik, Technikerstr. 25/8, 6020 Innsbruck, Austria}

   \date{\today}

\authorrunning{K.\ Tanidis et al.}

 
  \abstract
   {The {cosmological surveys that are} planned for the current decade will provide us with unparalleled observations of the distribution of galaxies on cosmic scales, by means of which we can probe the underlying large-scale structure (LSS) of the Universe. This will allow us to test the concordance cosmological model and its extensions. However, precision pushes us to high levels of accuracy in the theoretical modelling of the LSS observables, {so that no biases are introduced into} the estimation of {the} cosmological parameters. In particular, effects such as redshift-space distortions (RSD) can become relevant in the computation of harmonic-space power spectra even for the clustering of the photometrically selected galaxies, as {has previously been shown in literature.}}
   {In this work, we investigate the contribution of linear RSD, as formulated in the Limber approximation by a previous work, in forecast cosmological analyses with the photometric galaxy sample of the \Euclid survey. {We aim} to assess their impact and {to} quantify the bias on the measurement of cosmological parameters that {would be caused if this effect were neglected.}}
   {We {performed} this task by producing mock power spectra for photometric galaxy clustering and weak lensing, {as is} expected to be obtained from the \Euclid survey. We then {used} a Markov chain Monte Carlo approach to obtain the posterior distributions of cosmological parameters from {these} simulated observations.}
   {{When} the linear RSD {is neglected, significant biases are caused when galaxy correlations are used} alone and when {they} are combined with cosmic shear in the so-called 3$\times$2pt approach. {These} biases can be {equivalent to as much} as $5\,\sigma$ {when} an underlying $\Lambda$CDM cosmology \text{is assumed}. {When} the cosmological model {is extended} to include the equation-of-state parameters of dark energy, the extension parameters can be shifted by more than $1\,\sigma$.} 
   {}


   \maketitle
%

\section{Introduction}\label{sec:intro}

Within the current decade, several large-scale structure (LSS) surveys are expected to start their operations or to release their first results. {They} will provide a significant improvement to available cosmological data sets. {These} forthcoming LSS surveys will map the matter distribution in the Universe with exquisite precision. Some of {the surveys} will be ground-based, such as the Dark Energy Spectroscopic Instrument 
\citep[DESI,][]{2016arXiv161100036D,2016arXiv161100037D}, the 
Legacy Survey of Space and Time (LSST) at the Vera C.\ Rubin Observatory \citep{2009arXiv0912.0201L,2018arXiv180901669T,Ivezic:2008fe}, and the {Square Kilometre Array Observatory} \citep[SKAO; see, e.g.,][]{2015aska.confE..17A,2015aska.confE..19S,2015aska.confE..23B,2015aska.confE..24B,2015aska.confE..25C,2015aska.confE..31R,2020PASA...37....7S}. Others will be space-borne, such as the \Euclid satellite \citep{Laureijs2011,Amendola2013,Amendola2016,ISTF}, the \textit{Nancy Grace Roman} Space Telescope \citep{2015arXiv150303757S}, and the Spectro-Photometer for the History of the Universe, Epoch of Reionization, and Ices Explorer \citep[SPHEREx; see, e.g.,][]{2014arXiv1412.4872D,2018arXiv180505489D}. 

All {these} surveys rely on the observation of galaxy positions and shapes, with which {summary statistics can be constructed that are} customarily referred to in cosmology as galaxy clustering (GC) and weak lensing (WL). This can be done in multiple ways, {using either} the real-space 2pt correlation function, or the harmonic-space power spectrum {that} we study here, or even via other statistics {such as} COSEBIs \citep{2010A&A...520A.116S}, or higher-order statistics \citep{2023arXiv230112890E}. The GC encodes information on the clustering of matter due to the effect of gravity, while the WL {provides information} on the projected matter distribution {through} its gravitational lensing effect.

In this work, we focus on \Euclid and its surveys.\footnote{\url{http://www.euclid-ec.org/}.} \Euclid is a European Space Agency medium-class space mission whose launch took place on {1} July 2023. It will perform photometric and spectroscopic galaxy surveys over an area of $\sim15000\,\deg^2$ of the extragalactic sky \citep{Laureijs2011}, {with} the near-infrared instrument \citep{NISP} and the visible imager \citep{VIS}, which will be carried on board. The photometric survey will measure the positions and shapes of over a billion galaxies, enabling the analysis of photometric GC (GCph) and WL. {Because the} photometric measurements will provide relatively uncertain redshift measurements (compared to spectroscopic observations), the analyses of these observables will be performed via a tomographic approach {by} binning galaxies in redshift slices and considering the projected two-dimensional data sets. The precise radial measurements of the spectroscopic survey will instead allow us to perform a spectroscopic GC (GCsp) analysis, that is, {a galaxy-clustering} analysis in three dimensions.

In \citet[][`Euclid preparation: VII', hereafter EP:VII]{ISTF}, the constraints expected from \Euclid have been forecast for the individual GCsp, GCph, and WL probes and {also for their combination. To obtain} these results, \citetalias{ISTF} used several assumptions to simplify the theoretical computation of observables: the Limber approximation was used for all the photometric observables, and it was assumed {for GCph} that the only non-negligible contribution to the galaxy {position} correlation function comes from the anisotropies in the density field. It is known, however, that several other effects contribute to GCph, including lensing magnification, velocity, and relativistic effects \citep{Yoo2010,ChallinorandLewis2011,BonvinDurrer2011}.

{These} various contributions can be significant at very large scales, which we define as {scales} corresponding to a wavenumber smaller than that at which the matter power spectrum {peaks} at the matter-radiation equality scale. {These} scales are effectively within reach of wide surveys such as \Euclid, and it has been shown that neglecting them could lead to inaccurate results. {The cosmology that is} recovered through parameter estimation pipelines {might be} significantly biased with respect to the true underlying cosmology \citep{2015MNRAS.451L..80C,Tanidis:2019teo,Tanidis_mag,Martinelli:2021ahc,Lepori2021}.

In this work, we focus on one of the most important effects, namely redshift-space distortion (RSD). {We also aim specifically to quantify} its impact on the expected results of the \Euclid wide survey. The effect of the linear RSD on the angular clustering is not new and has been thoroughly studied {before} \citep{Fisher1993, Heaven1995, Padmanabhan, Blake, Nock,Crocce, Balaguera-Antolinez2018,Tanidis:2019teo}. {The effects} have also been included in the Dark Energy Survey \citep[DES,][]{thedarkenergysurveycollaboration2005dark} 3$\times$2pt data analysis for Y3 in the configuration space \citep{Abbott21}, {although} they were initially neglected in the Y1 analysis for the GC in the configuration and harmonic space \citep{Elvin-Poole:2017xsf, Andrade-Oliveira2021}. Here, we apply the approach of \citet{Tanidis:2019teo}, where the linear RSD contribution to GC harmonic-space power spectra is obtained within the Limber approximation. {We examine this contribution to} the \Euclid wide survey. This allows us to {compute the} theoretical prediction at a reasonable speed {so that it can be used to estimate the parameters.} Moreover, {we only} focus on GCph {and do not} discuss GCsp at all. {We therefore} always refer to the GCph probe simply as GC {throughout}.

The paper is organised as follows. The equations {we} used to compute the theoretical predictions for the observables of interest are reviewed in \Cref{sec:theory}, where we also outline how RSD {enters} the calculations. In \Cref{sec:flagship} we summarise the results of the Flagship simulation galaxy catalogue. Our analysis method is shown in \Cref{sec:method} and the results {we obtained are provided} in \Cref{sec:results}. We finally summarise our conclusions in \Cref{sec:conclusions}.

\section{Photometric observables in \textit{Euclid}}\label{sec:theory}

\subsection{Harmonic-space power spectra}
\label{sec:harmonic-space}
The harmonic-space power spectrum $C_{\ell}^{AB}(z_i,z_j)$ between an observable $A$ in the redshift bin $i$ and an observable $B$ in the redshift bin $j$ is defined as

\begin{equation}
    \left\langle A_{i,\ell m}\,B^\ast_{j,\ell' m'}\right\rangle=C^{AB}_\ell(z_i,z_j)\,\delta^{\rm K}_{\ell\ell'}\,\delta^{\rm K}_{m m'}\;,
    \label{eq:Harmonic_definition}
\end{equation}
where $X_{\ell m}$ are the coefficients of the harmonic expansion of observable $X$, and $\delta^{\rm K}$ denotes the Kronecker symbol. Here, {the} letters $A$ and $B$ stand for the observables of interest: galaxy number {count} fluctuations, $\Delta$, or galaxy ellipticities, $\epsilon$.
Both fields are discussed in more detail in \Cref{sec:Limber_RSDs} and \Cref{sec:Shear}, respectively. 

Within the Limber approximation \citep{Kaiser1992}, {which is} valid at $\ell\gg 1$ and for broad redshift kernels, the harmonic-space (also {called} angular) power spectrum between two observables $A$ and $B$ is
\begin{equation}
C^{AB}_{ij}(\ell)=\int \frac{\de r}{r^2}\, W_i^A(\ell,r)\,W_j^B(\ell, r)\,P_{\delta\delta}\left(k=\frac{\ell+1/2}{r},r\right)\;,
    \label{eq:Harmonic_limber}
\end{equation}
where $P_{\delta\delta}$ is the (non-linear) matter power spectrum, $k=|\vec k|$ {is} the wave number, which is the Fourier mode related to the comoving separation between pairs of galaxies in configuration space, and $r(z)$ is the radial comoving distance to redshift $z$ for a flat cosmology. The generic redshift-binned kernel $W_i^A(\ell,r)$ takes different forms depending on the target observables{, as we show in} \Cref{sec:Limber_RSDs} and \Cref{sec:Shear}. {We} use the notation $C^{AB}_{ij}(\ell)$ as in \citetalias{ISTF}, as opposed to $C^{AB}_\ell(z_i,z_j)$ of \Cref{eq:Harmonic_definition}, to denote the fact that we refer to the Limber-approximated power spectrum.

\subsection{Linear RSD in GCph}
\label{sec:Limber_RSDs}
In GC, galaxies are biased tracers of the underlying matter field \citep{Kaiser1987}. At sufficiently large scales, the bias can be considered {to only depend} on redshift and not on scale \citep{Abbott:2018ydy}. On the other hand, when non-linear scales {are added}, the galaxy bias becomes non-local and a specific treatment {is required to} account for this effect \citep{10.1093/mnras/stw2443, Desjacques_2018}. {In addition, RSDs} are additional observational effects due to the peculiar velocities of galaxies \citep{Kaiser1987,Szalay1998}. These are customarily split into linear RSD ({also} known as\ the Kaiser effect) and non-linear RSD ({also known as the fingers of God, `FoG'} hereafter). The former {causes} the squashing of the galaxy 2pt correlation function on large scales in the direction perpendicular to the line of sight, while the latter enhances the clustering amplitude along the line-of-sight direction on small scales. In the current analysis we consider a simple model, {that assumes that} the galaxy bias is linear and {scale independent}, and we only account for linear RSD. The modelling of the non-linear galaxy bias and the FoG as well as their effects in the photometric observables for \textit{Euclid} is left for future work. The kernel of \Cref{eq:Harmonic_limber} {for the GC that includes} up to linear RSD in the Limber approximation \citep[for details, see][]{Tanidis:2019teo} takes the form
\begin{equation}
W_i^\Delta(\ell,r)=W_i^{\rm den}(r)+W_i^{\rm RSD}(\ell,r)\;,
\label{eq:GC_kernel}
\end{equation}
with the first term being due to fluctuations in the density field,
\begin{equation}
W_i^{\rm den}(r)=n_i(r)\,b_i(r)\;,
    \label{eq:den}
\end{equation}
with $n_i(r)\,\de r$ the galaxy probability density in bin $i$ between {the} comoving distance $r$ and $r+\de r$, and $b_i$ being the corresponding linear galaxy bias, computed at $r\equiv r(z)$. {We treat this} as a constant within the redshift bin, and its actual amplitude is a nuisance parameter, {over which we marginalised} in our analysis. For the fiducial galaxy bias values in each bin (see \Cref{table:fiducials2}), {we used} the fiducial model of \citet[][`Euclid Colaboration: XII', hereafter EP:XII]{Pocino2021}{, who} considered a magnitude cut at $\IE=24.5$ for the \textit{Euclid} imager{, which} will observe through an optical broad band. For the galaxy distributions $n_i(r)$, we {used} the outcome of the Flagship galaxy simulation of the Euclid Consortium, for which we provide details in \Cref{sec:flagship}.

The second term  in \Cref{eq:GC_kernel} is the RSD contribution \citep{Tanidis:2019teo},
\begin{equation}
W_i^{\rm RSD}(\ell,r)=\sum_{n=-1}^1 L_{n}(\ell)\,n_i\left(\frac{2\,\ell+1+4\,n}{2\,\ell+1}\,r\right)\,f\left(\frac{2\,\ell+1+4\,n}{2\,\ell+1}\,r\right)\;,
\label{eq:rsd}
\end{equation}
with
\begin{align}
L_{0}(\ell)&=\frac{2\,\ell^2+2\,\ell-1}{(2\,\ell-1)\,(2\,\ell+3)}\;,\\
L_{-1}(\ell)&=-\frac{\ell\,(\ell-1)}{(2\,\ell-1)\sqrt{(2\,\ell-3)\,(2\,\ell+1)}}\;,\\
L_{+1}(\ell)&=-\frac{(\ell+1)\,(\ell+2)}{(2\,\ell+3)\sqrt{(2\,\ell+1)\,(2\,\ell+5)}}\;.
\label{eq:Ls}
\end{align}

\begin{figure}
    \centering
    \includegraphics[width=\columnwidth]{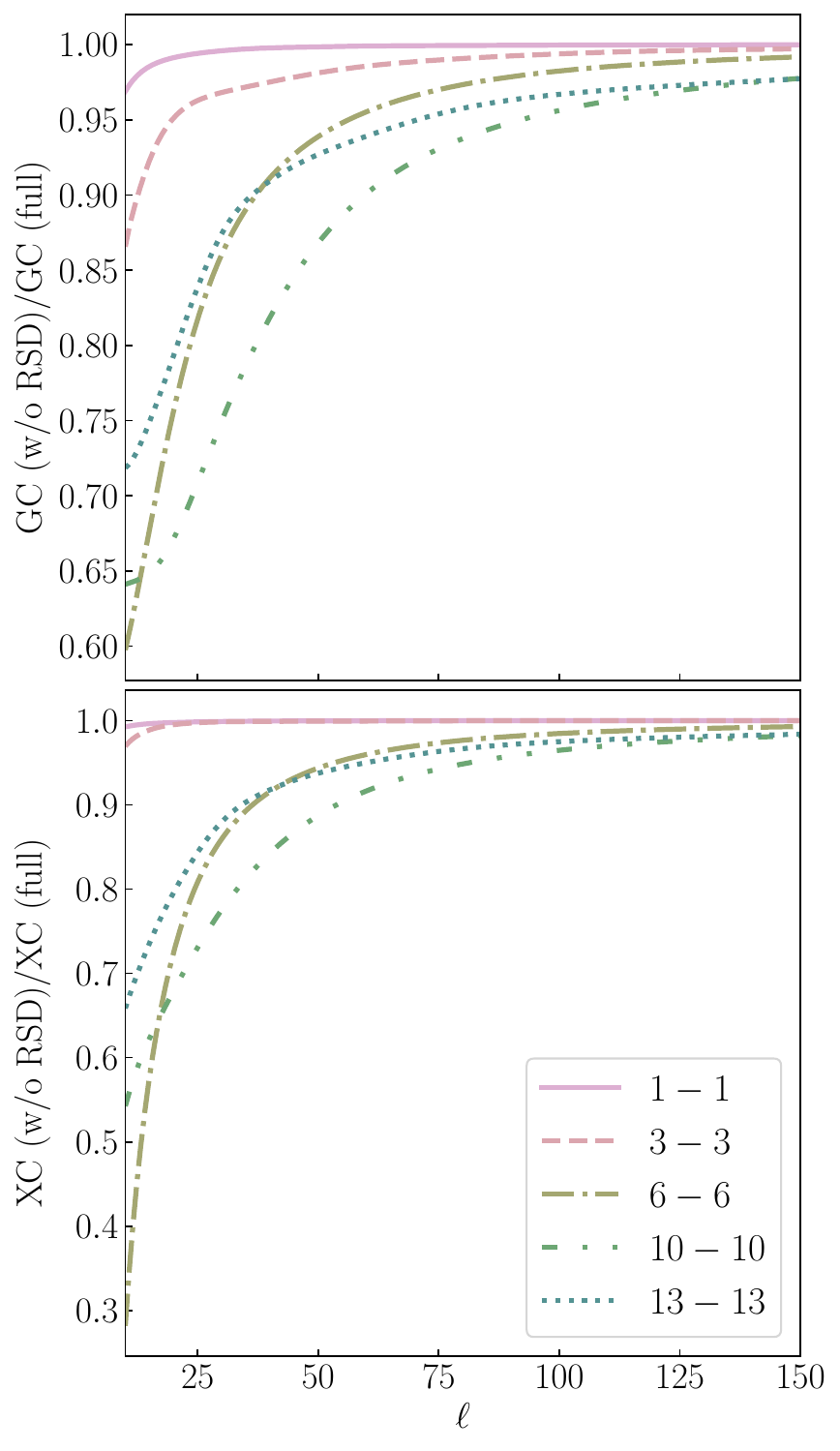}
    \caption{Ratio of the harmonic-space power spectrum for the density fluctuations alone with respect to that including RSD for some tomographic auto-bin correlations ($i=j$). \textit{Top:} GC. \textit{Bottom:} XC.}
    \label{fig:RSD_bins}
\end{figure}

In \Cref{eq:rsd}, $f\coloneqq-(1+z)\,\de\ln D/\de z$ is the growth rate, with $D$ being
the linear growth factor. In the top panel of \Cref{fig:RSD_bins}, we quantify the signal loss when the linear RSD are not included in the photometric GC harmonic-space power spectrum. In particular, RSD are important at large scales, {for approximately} $\ell\lesssim100$. Additionally, the contribution to the total signal increases with redshift for a given redshift bin width. For example, RSD contribute to the total signal from 2--3\% at low redshift ({solid purple} curve, bin pair 1--1) and gradually increase with increasing redshift (dash-dotted {yellow} curve, bin pair 6--6) up to $\sim40\%$ at the lowest available multipoles. However, we should note that this is not true for the highest redshift bin ({dotted blue} curve, bin pair 13--13). {This} bin contributes less to the full signal than the bin pairs 10--10 and 6--6{, for example, \citet{Tanidis_mag} showed} that the linear RSD effect is gradually diluted when the width of the redshift bin increases. This is particularly the case for the {highest-redshift} bins due to the large photometric uncertainties at high redshifts.

Another important correction to the galaxy density field is the magnification bias. {We wish} to quantify the impact of linear RSD alone on \Euclid photometric observables {here and therefore neglected} the magnification effect in our analysis. However, this effect has been thoroughly studied in \citet{Lepori2021} (in that study the linear RSD were neglected), and its inclusion {was} found to be crucial to avoid biases on the cosmological parameter estimation for \textit{Euclid}. Similar studies {of} this effect have also been conducted for other future experiments \citep{Tanidis_mag}.

{In addition,} there are also local and integrated contributions to the signal that are measurable at ultra-large scales, such as the Doppler terms, the Sachs-Wolfe and the integrated Sachs-Wolfe effects, and the time delay. We {neglected} these contributions in our analysis {because} for $\ell\gg1$, where the Limber approximation holds, their effect is negligible \citep{Yoo2010,ChallinorandLewis2011,BonvinDurrer2011,Martinelli:2021ahc}.

\subsection{Cosmic shear}
\label{sec:Shear}

The LSS of the Universe deflects the paths of photons {that are} emitted by distant sources. This distorts the source images. {This} distortion is decomposed into the convergence, $\kappa$, and shear, $\gamma$, which correspond to size magnification and shape distortion of the images, {respectively, and are related linearly}. Both signals contain useful cosmological information, but the former is more impossible to extract because it requires knowledge of the original source sizes \citep{2013MNRAS.433L...6H,Alsing2015}. For this reason, shear is the usual focus of {WL} surveys of the LSS. 

The harmonic-space power spectrum of the shear field is a probe of the growth of structures and the cosmological expansion. {In addition to} the cosmic shear, the correlation of the galaxy shapes also receives a contribution from intrinsic alignments (hereafter IA), which in the context of cosmological {WL} studies is regarded as a systematic effect. This accounts for the fact that, {in addition to} the random orientations of galaxies, {which ideally would} make the ellipticity an unbiased estimator of the shear field, there are also IA of the galaxies {that are caused by} the tidal interactions during galaxy formation, {and also astrophysical effects that contaminate the WL} analysis \citep{Joachimi2015}.

For the WL sample, the kernel of {the} observed ellipticity power spectrum including cosmic shear $\gamma$ and IA, reads
\begin{equation}
W^\epsilon_i(\ell,r)=W^{\gamma}_i(\ell,r)+W^{\rm IA}_i(r)\;.
    \label{eq:WL_kernel}
\end{equation}
The shear contribution is
\begin{equation}
W^{\gamma}_i(\ell,r) = \frac{3\,L_\gamma (\ell)\,\om\,H_0^2}{2c^2}\left[1+z(r)\right]\,r\,q_i(r)\;,
    \label{eq:shear}
\end{equation}
where $c$ is the vacuum speed of light, $q_i(r)$ is the so-called lensing efficiency for a flat Universe,
\begin{equation}
q_i(r)=\int_{r}^{\infty}{\rm d}r'\,\frac{r'-r}{r'}\,n_i(r')\;,
    \label{eq:lens_eff}
\end{equation}
and the $\ell$-dependent factor is given by
\begin{equation}
L_\gamma(\ell)=\sqrt{\frac{(\ell+2)!}{(\ell-2)!}}\left(\frac{2}{2\,\ell+1}\right)^2\;.
    \label{eq:len_factor}
\end{equation}
For WL {we also used} the $n_i(r)$ obtained from the Flagship simulation of \Cref{sec:flagship}. {Because we restricted our analysis to multipoles, for which} the Limber approximation applies ($\ell\gg1$), this factor can be considered {to be} $L_\gamma(\ell)\approx1$.

The IA contribution can instead be modelled as in \citetalias{ISTF}, namely
\begin{equation}
W^{\rm IA}_i(r)=-\mathcal A_{\rm IA} \,\mathcal C_{\rm IA}\,\om\, \frac{\mathcal F_{\rm IA}\left[z(r)\right]}{D\left[z(r)\right]}\,n_i(r)\;,
    \label{eq:in_al}
\end{equation}
where
\begin{equation}
\mathcal F_{\rm IA}(z)= (1+z)^{\eta_{\rm IA}}\, \left[\frac{\langle L \rangle (z)}{L_{\ast}(z)}\right]^{\beta_{\rm IA}}\;.
    \label{eq:lum_factor}
\end{equation}
The amplitude and shape of the IA signal is captured by the nuisance parameters {$\mathcal A_{\rm IA}$, $\beta_{\rm IA}$ and $\eta_{\rm IA}$} (see \Cref{table:fiducials2} for their fiducial values), with $C_{\rm IA}$ kept fixed at the value $0.0134$ {because} it is degenerate with $\mathcal A_{\rm IA}$. The terms $\langle L \rangle(z)$ and  $L_{\ast}(z)$ denote the mean and the characteristic luminosity of the source galaxies with respect to redshift.\footnote{We refer the reader to \citetalias{ISTF} for more details concerning the $\text{IA}$ modelling.} {We neglected} other sources of systematic effects for the WL probe such as the shear bias and the photometric redshift uncertainties ({which} are also present for the tomographic bins in photometric GC in {principle}).

\begin{figure}
    \centering
    \includegraphics[width=\columnwidth]{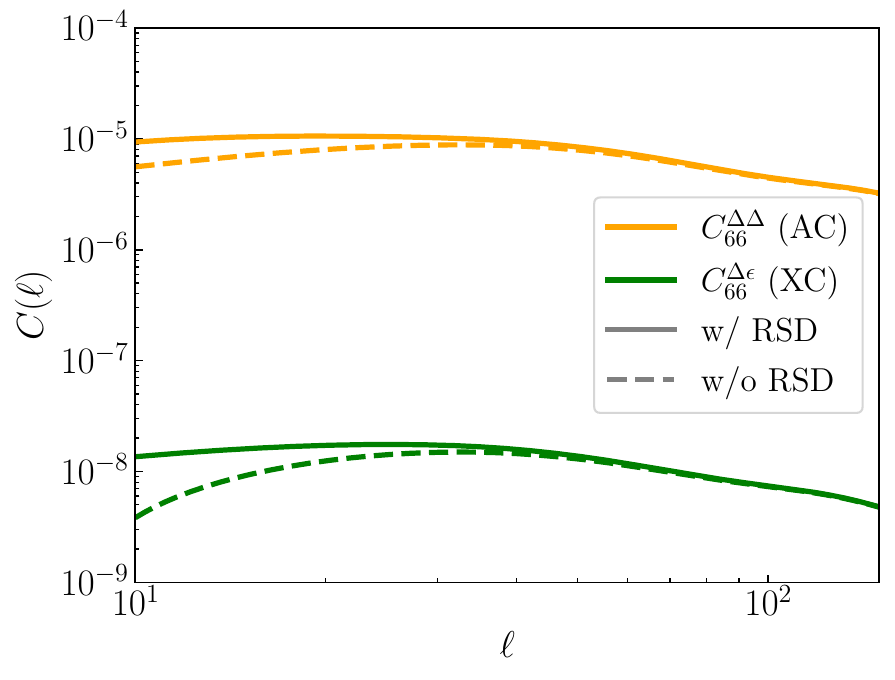}
    \caption{GC (orange) and the XC (green) spectrum for the bin correlation 6--6. {The dashed} lines correspond {to} spectra {without} RSD{, and the} solid {lines show spectra with RSD.}}
    \label{fig:spectra}
\end{figure}

In the bottom panel of \Cref{fig:RSD_bins} we show the signal loss {when} we neglect the linear RSD contribution in the cross power spectrum between GC and WL (hereafter XC). The picture is very similar compared to the one presented for the GC (top panel of \Cref{fig:RSD_bins}). The main difference is that the RSD contribution to the total signal is {lower} than $\sim1\%$ at the lowest redshift (bin pair 1--1), reaching a maximum of up to $\sim70\%$ in bin pair 6--6 {at the largest scales}. 

{For} some bin cases{, the signal that is lost when the RSD is neglected} with respect to the total signal in XC is more than the GC, which indeed seems to be counter-intuitive {because} the RSD is an additional term in the GC kernel \Cref{eq:GC_kernel}. Although it is true that the RSD kernel given by \Cref{eq:rsd} appears twice in \Cref{eq:Harmonic_limber} for all bin correlations due to \Cref{eq:GC_kernel} and only once in the XC, the GC spectra have more power than the XC {spectra, as shown in \Cref{fig:spectra}}. The WL kernel of \Cref{eq:WL_kernel} also appears once in the XC{, and this kernel lowers the} signal. For this reason, the XC spectra {might} be numerically more {strongly affected when the} RSD signal {stronger, as} in the bin correlations 6--6 and 10--10.
 
\subsection{The \Euclid Flagship simulation}\label{sec:flagship}
In order to create realistic mock data vectors for our analysis, we {used} the simulated results from the Flagship galaxy simulation of the Euclid Consortium (Euclid Consortium, in preparation). The galaxy catalogue was produced using the $N$-body Flagship dark matter simulation \citep{Potter2017} with a $\Lambda$CDM fiducial cosmology given by the total matter abundance, $\om=0.319$; the baryon abundance, $\ob=0.049$; the r.m.s.\ variance of the linear matter fluctuations at $z=0$ in spheres {with a radius of} $8\,h^{-1}\,\mathrm{Mpc}$, $\sigma_8=0.830$; the spectral index of the primordial curvature power spectrum, $n_{\rm s}=0.96$; and the dimensionless Hubble constant, $h\equiv H_0/(100\,\mathrm{km\,s^{-1}\,Mpc^{-1}})=0.67$. The $N$-body simulation ran a box of $3.78\,h^{-1}\,\mathrm{Gpc}$ with a particle mass of $2.398\times10^9\,h^{-1}\,\mathit{M}_{\odot}$.

The dark matter haloes were identified using \texttt{ROCKSTAR} \citep{Rockstar} down to masses of $2.4\times10^{10}\,h^{-1}\,\mathit{M}_{\odot}$ (corresponding to ten particles per halo). Then, the galaxies were assigned to the haloes using the {halo-occupation} distribution and the {halo-abundance} matching methods, following the recipe presented in \citet{Carretero2015}. Several observational constraints were used to calibrate the galaxy mocks including the luminosity function \citep{Blanton2003,Blanton2005a} applied for the faint galaxies, the measurements of galaxy clustering as a function of colour and luminosity \citep{Zehavi2011}, {and} the colour-magnitude diagram  from \citet{Blanton2005b}. The final galaxy catalogue contains {almost} $3.4$ billion galaxies over $5000\,$deg$^2$ and extends up to redshift $2.3$.

\begin{figure}
    \centering
    \includegraphics[width=\columnwidth]{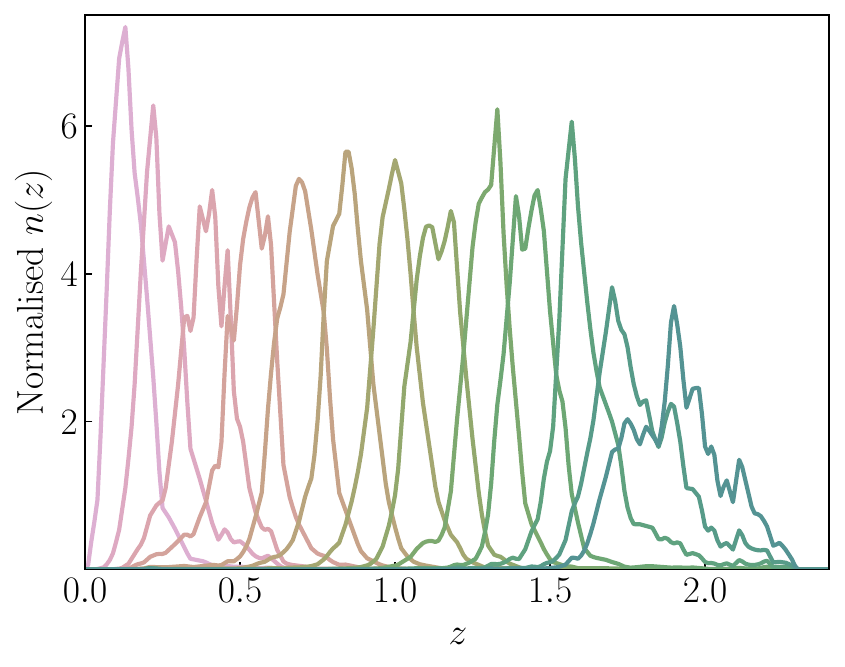}
    \caption{{Normalised} number densities of the $13$ tomographic {equispaced} bins with redshift from the \textit{Euclid} Flagship simulation \citepalias{Pocino2021}.}
    \label{fig:n_of_z}
\end{figure}

{We followed} the analysis of \citetalias{Pocino2021}, where an optimisation of the galaxy sample for photometric GC analyses was performed using the Flagship simulation. \citetalias{Pocino2021} generated photometric redshift estimates with the {directional-neighbourhood-fitting} training-based algorithm \citep{DNF} for all galaxies within a patch of $400\,\deg^2$ of the Flagship simulation up to a magnitude limit of $25$ in the VIS band. {We considered} the fiducial sample from \citetalias{Pocino2021}. This corresponds to a training of the algorithm with an incomplete spectroscopic training sample to mimic the lack of spectroscopic information at very faint magnitudes, and to optimistic magnitude limits for all photometric bands. There is an additional selection of objects with magnitudes brighter than $24.5$ in the VIS band. By using the {directional-neighbourhood-fitting} algorithm, two different estimates for the photometric redshift are provided for each object. One {estimate} is the average of the redshifts from the neighbourhood, {which} we denote $z_{\rm mean}$. The second estimate is a Monte Carlo draw from the nearest neighbour{, and we denote this} $z_{\rm mc}$. We refer to \citetalias{Pocino2021} and \citet{DNF} for the similarities and differences between these two estimates. The final sample {was} then composed of $13$ tomographic equispaced bins in $z_{\rm mean}$ up to $z=2$, that is, $13$ bins with {a} constant redshift width in $z_{\rm mean}${, and therefore, with a} different width in $z_{\rm mc}$. The normalised number densities of these bins as a function of redshift are shown in \Cref{fig:n_of_z}. In addition to the galaxy distributions, we also {considered} the linear galaxy biases for each of these distributions, which are provided in \citetalias{Pocino2021}.

\section{Synthetic data and analysis method}\label{sec:method}
In order to quantify the impact of linear RSD on {the  photometric observables of \Euclid, we createed mock data that we compared with} our theoretical predictions. For {this} purpose, we {computed} the power spectra for GC, WL, and XC following \Cref{eq:Harmonic_limber} in a fiducial cosmology. We {chose} this to be the same {as} \citetalias{ISTF}, namely a flat $\Lambda$CDM model with one massive and two massless neutrinos fixed to $\sum  m_\nu=0.06$eV. The fiducial values for the parameters are shown in \Cref{table:fiducials1} and \Cref{table:fiducials2}.

\begin{table}
\centering
\caption{Fiducial values of {the} cosmological parameters.}
\label{table:fiducials1}
\begin{tabularx}{\columnwidth}{XXXXXXX}
    \hline
    $\om$ & $\ob$ & $h$ & $\sigma_8$ & $n_{\rm s}$ & $w_0$ & $w_a$ \\
    \hline
    $0.320$ & $0.050$ & $0.67$ & $0.816$ & $0.96$ & $-1.0$ & $0.0$ \\
    \hline
\end{tabularx}
\end{table}

\begin{table*}
\centering
\caption{Fiducial values of {the} nuisance parameters.}
\label{table:fiducials2}
\begin{tabularx}{\textwidth}{XXXXXXXXXXXXXXXX}
    \hline
    $b_1$ & $b_2$ & $b_3$ & $b_4$ & $b_5$ & $b_6$ & $b_7$ & $b_8$ & $b_9$ & $b_{10}$ & $b_{11}$ & $b_{12}$ & $b_{13}$ & $\mathcal A_{\text{IA}}$ & $\beta_{\text{IA}}$ & $\eta_{\text{IA}}$ \\
    \hline
    $1.025$ & $1.037$ & $1.066$ & $1.110$ & $1.198$ & $1.293$ & $1.429$ & $1.559$ & $1.758$ & $1.947$ & $2.217$ & $2.537$ & $2.738$ & $1.72$ & $2.17$ & $-0.41$ \\
    \hline
    \smallskip
\end{tabularx}
{\small
\raggedright \textit{Note:} The linear galaxy bias parameter per redshift bin $i$ for GC is denoted with $b_i$ and the WL parameters for the whole redshift range with $\mathcal A_{\text{IA}}$ , $\beta_{\text{IA}}$, and $\eta_{\text{IA}}$ (for the resources of the IA modelling, see again \citetalias{ISTF}.). The centres of the redshift bins are the following: $\bar{z}_1=0.14$, $\bar{z}_2=0.26$, $\bar{z}_3=0.39$, $\bar{z}_4=0.53$, $\bar{z}_5=0.69$, $\bar{z}_6=0.84$, $\bar{z}_7=1.0$, $\bar{z}_8=1.14$, $\bar{z}_9=1.3$, $\bar{z}_{10}=1.44$, $\bar{z}_{11}=1.62$, $\bar{z}_{12}=1.78$ and $\bar{z}_{13}=1.72$. \par}
\end{table*}

We {modelled} the matter power spectrum, $P_{\delta\delta}$, on non-linear small scales using \texttt{halofit} \citep{Smith2003}, including corrections for both dark energy \citep{Takahashi2012} and massive neutrinos \citep{Bird2012} as in \citetalias{ISTF}. Using these assumptions, we can compute the harmonic-space power spectra of \Cref{eq:Harmonic_limber} for photometric GC, WL, and XC. These represent our synthetic data set, against which we can compare theoretical predictions from different models to constrain model parameters. In order to do {this, we obtained} the $\chi^2$ for any set of parameters $\vec{\theta}$ including cosmological and nuisance, and {assumed} the Gaussian likelihood
\begin{equation}
\chi^2(\vec\theta)=\left[\vec d-\vec t(\vec\theta)\right]^{\sf T} \tens C^{-1} \left[\vec d-\vec t(\vec\theta)\right]\;.
    \label{eq:chi2}
\end{equation}
{The} data vector $\vec d$, the vector of theoretical predictions $\vec t({\vec \theta})$, and the covariance matrix $\tens C$ (which is assumed to be constant and independent of the cosmological parameters), {were all} stacked along the $i$, $j$, and $\ell$ indices.

The covariance matrix of the data, $\tens C$, is the flattened version of the fourth-order Gaussian covariance (as in \citetalias{ISTF}), namely,
\begin{multline}
\label{eq:covmat}
\text{Cov}\left[C^{AB}_{ij}(\ell),C^{A'B'}_{mn}(\ell')\right]=\frac{\widetilde C^{AA'}_{im}(\ell)\,\widetilde C^{BB'}_{jn}(\ell)+\widetilde C^{AB'}_{in}(\ell)\,\widetilde C^{BA'}_{jm}(\ell)}{(2\,\ell+1)\,\Delta\ell\,f_{\rm sky}}\,\delta^{\rm K}_{\ell\ell'}\;,
\end{multline}
with $\widetilde C^{AB}_{ij}(\ell)=C^{AB}_{ij}(\ell)+N^{AB}_{ij}(\ell)$, and $N^{AB}_{ij}(\ell)$ being the noise contribution to the measurement, and $f_{\rm sky}=0.3636$ is the sky fraction observed by \textit{Euclid}. Finally, we {used}
\begin{align}
    N^{\Delta\Delta}_{ij}(\ell) &=  \frac1{\bar n_i}\,\delta^{\rm K}_{ij}\;,\\
    N^{\epsilon\epsilon}_{ij}(\ell)&= \frac{\sigma^2_\epsilon}{\bar n_i}\,\delta^{\rm K}_{ij}\;,\\
    N^{\Delta\epsilon}_{ij}(\ell)&= 0\;,
\end{align}
with $\bar n_i$ being the number density of galaxies per steradian in each tomographic bin and $\sigma_\epsilon^2$ the intrinsic ellipticity variance, which we {assumed} to be $\sigma_\epsilon=0.3$ \citepalias[as in][]{ISTF}.

We {binned} our data in multipoles considering $N_\ell=20$ log-spaced multipole bins in the range $\{\ell_{\rm min}, \ell_{\rm max}\}$, with $\Delta \ell$ the width of each bin. We {considered} the lowest multipole to be $\ell_{\rm min}=10$ and {considered} an optimistic and a pessimistic scenario for the maximum multipole cut $\ell_{\rm max}$ \citepalias[see][]{ISTF}, with
\begin{itemize}
    \item Optimistic | pessimistic GC: 3000 | 750,
    \item Optimistic | pessimistic WL: 5000 | 1500.
\end{itemize}
For XC, we conservatively considered the smallest $\ell_{\rm max}$ (corresponding to the GC values).

{The assumptions we adopted} from \citetalias{ISTF} for the purposes of this work in our modelling are not expected to {affect the results strongly}. For example, an increase in $\sigma_\epsilon$ would have an effect on the WL part but not on the GC and XC. Then, a decrease {in} ${\bar n_i}$ would increase the shot noise, but the GC part of the analysis, which is {what} we are most interested in, with the linear RSD scales, should not be {dominated by shot noise}. Therefore, the 3$\times$2pt contours would increase through the WL contribution, {but would not change} our conclusions on RSD.

With this approach, we {needed only one additional} ingredient to obtain the $\chi^2$ of \Cref{eq:chi2}, which is the theory vector $\vec t(\vec \theta)$. We {chose} to focus on two models to be constrained with our mock data, which we refer to as $\Lambda$CDM and $w_0w_a$CDM flat cosmologies following the same modelling as in \citetalias{ISTF}. The former has {five} free cosmological parameters{, which are} $\om$,  $\ob$, $h$, $\sigma_8$, and $n_{\rm s}$. The latter also includes the equation-of-state parameters of dark energy, $w_0$ and $w_a$, which come from the CPL {parametrisation} \citep{doi:10.1142/S0218271801000822,PhysRevLett.90.091301},
\begin{equation}
w(z)=w_0+w_a\frac{z}{1+z}\;.
    \label{eq:CPL}
\end{equation}
As nuisance parameters, both models feature {for the GC, the} linear galaxy bias amplitudes $b_i$ (see \Cref{sec:Limber_RSDs}){, and for the WL, the} IA modelling parameters $\mathcal A_{\rm IA}$, $\beta_{\rm IA}$, and $\eta_{\rm IA}$ (see \Cref{sec:Shear}).

\section{Results}
\label{sec:results}
In this section, we use the method presented in \Cref{sec:method} to investigate the impact of the RSD contribution, modelled as in \Cref{sec:Limber_RSDs}, on the final cosmological constraints. We explore the parameter space with a Markov chain Monte Carlo (MCMC) approach using \emcee\ \citep{ForemanMackey2013}, and we obtain from \camb\ \citep{LCL2000,2012JCAP...04..027H} all the quantities needed to compute the theoretical predictions following \Cref{eq:Harmonic_limber} in a modified version of the code \cosmosis\ \citep{Zuntz2015}. We adopt improper (flat) priors for all the parameters in \Cref{table:fiducials1} and \Cref{table:fiducials2}.

\subsection{Validation}
\label{sec:Fisher_and_MCMC}
Before we {proceeded to set} up the realistic and computationally expensive MCMC chains for all the different cosmologies and scenarios, we first {ran} a single case with {an MCMC and compared} it to a Fisher matrix forecast \citepalias[see][for details about Fisher forecasts and their derivatives accuracy]{ISTF} in order to validate the pipeline. In brief, entries of the Fisher matrix are constructed as
\begin{equation}
F_{\alpha\beta}= \frac{\partial \vec t^{\sf T}}{\partial \theta_\alpha} \tens C^{-1} \frac{\partial \vec t}{\partial \theta_\beta}\;, 
    \label{eq:fisher}
\end{equation}
where $\{\theta_\alpha\}$ are the elements of the parameter vector $\vec \theta$ (see \Cref{sec:method}).

In general, the Fisher forecasts, whose computation is usually faster than any Bayesian sampler, {are} used in order to investigate the likelihood curvature of the parameter hyperspace near its peak. {When} the posterior is very well described by a Gaussian, the Fisher forecasts are particularly accurate, {and the smaller the uncertainties, the closer the peak of the posterior where the Gaussian approximation holds.}
This applies to forthcoming LSS experiments {such as} \Euclid, which will provide us with an {unprecedentedly} large number of sources that will minimise the resulting errors. As expected, the precision will increase even further when we consider all the available signal. Therefore, we {performed} a Fisher forecast on the combination of GC, WL, and XC, which we {labelled} 3$\times$2pt. We {show} this by considering a $w_0w_a$CDM cosmological model and an optimistic scale cut, in order to investigate the potential Gaussianity of the posterior in an extended cosmology model with two {additional} parameters, $\{w_0,w_a\}$. In \Cref{eq:fisher} we {calculated} the theory vector $\vec t$ for the fiducial cosmology model and the covariance matrix $\tens C$ following \Cref{sec:method}. We also {included} linear RSD as described in \Cref{sec:Limber_RSDs}. With the same assumptions, we {repeated} the analysis with an MCMC approach, exploring the posterior and then finding the minimum $\chi^2$ over the parameter space $\vec\theta$.

\begin{figure*}
    \centering
    \includegraphics[width=0.63\textwidth]{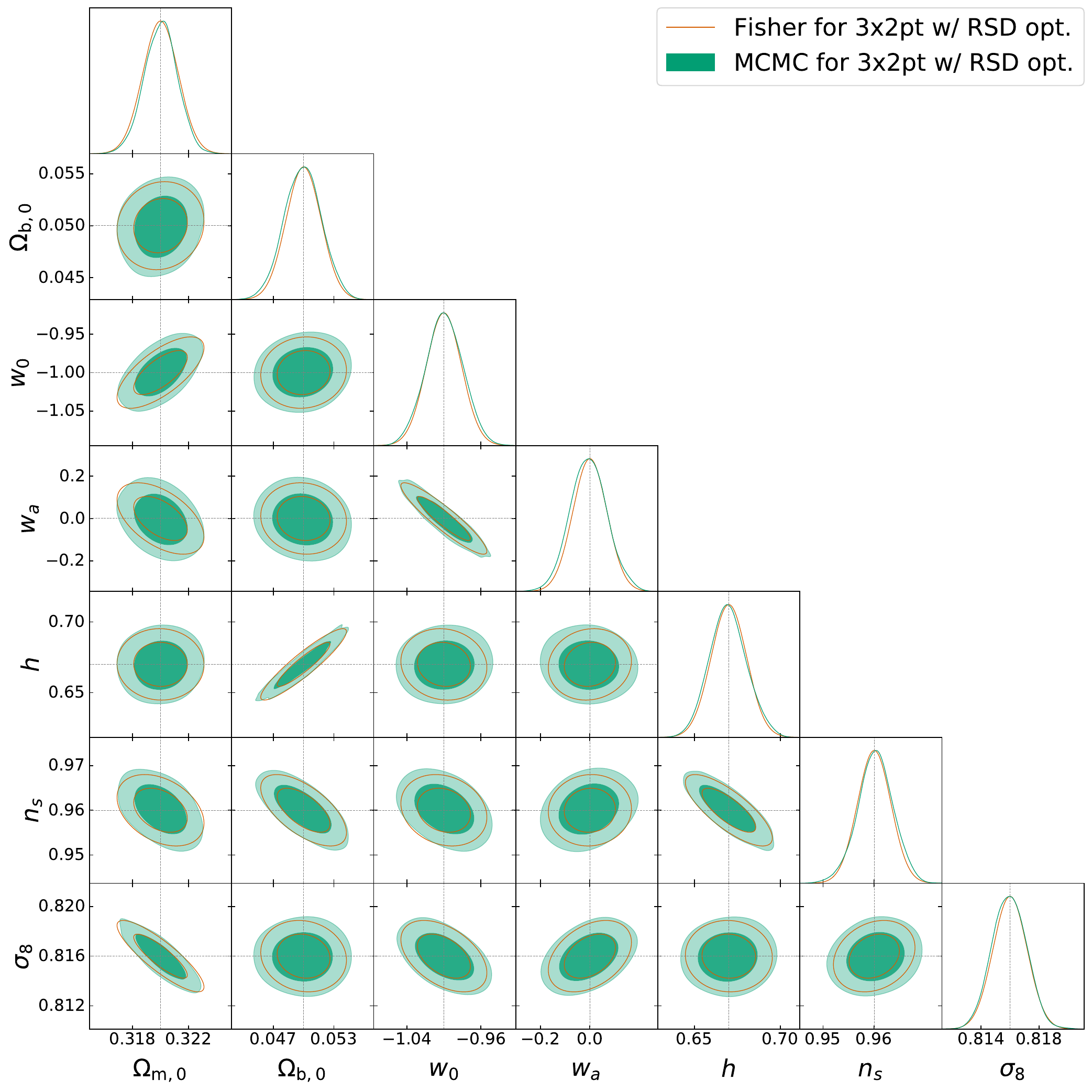}\\
    \includegraphics[width=0.63\textwidth]{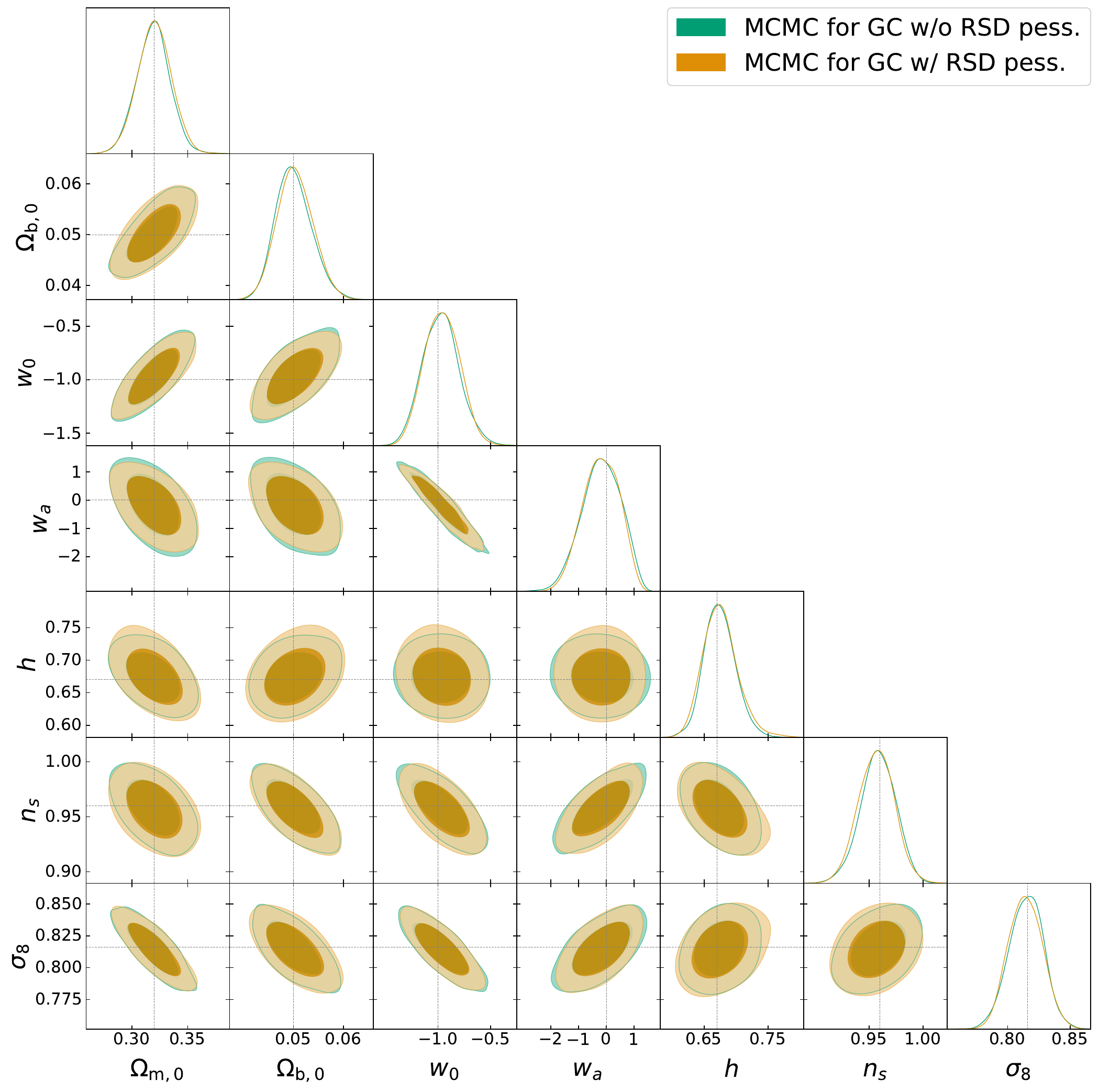}
    \caption{$68\%$ and $95\%$ C.L.\ marginalised contours along with the one-dimensional posterior distributions on the cosmological parameters. \textit{Top:} Constraints from the 3$\times$2pt analysis of the flat $w_0w_a$CDM model for the optimistic scale cut considering RSD. The filled green contours are the MCMC constraints, {and} the open orange {contours show} the Fisher ellipses for the same modelling. The fiducial cosmology model is marked with dotted black lines. \textit{Bottom:} Constraints from the GC alone of the flat $w_0w_a$CDM model (pessimistic scenario), {with} (orange) and {without} (green) RSD. The fiducial cosmology model is marked with dotted  black lines.}
    \label{fig:Fisher_MCMC_constpow}
\end{figure*}

The comparison between the MCMC and Fisher approaches is shown in the top panel of \Cref{fig:Fisher_MCMC_constpow}, where the $68\%$ and $95\%$ credibility levels (C.L.) on cosmological parameters are presented, after the bias and IA parameters {were} marginalised over.\footnote{{we note} that $68\%$ and $95\%$ C.L.\ exactly {correspond} to {one} and {two} standard deviations in the Gaussian approximation of the Fisher matrix after marginalising over the {remaining} parameters.}. The filled green contours correspond to the MCMC and the empty orange contours {show} the Fisher forecast. The agreement between the two is remarkable {and highlights} the Gaussianity of the posterior, with almost perfectly overlapping constraints and all the directions and widths of the Fisher ellipses recovered by the MCMC. In addition, the fiducial cosmology model values are presented with dotted {black lines. They are} located well within the $68\%$ C.L.\ contours.

\subsection{The contribution of RSD to constraining power}
{After we validated} our MCMC pipeline{, we investigated} how the additional information encoded in RSD affects the constraining power on the model parameters. {To do} this, we {considered GC alone} and not the total 3$\times$2pt {because} as shown in \Cref{sec:Limber_RSDs}, RSD is a correction for the GC signal alone{, and therefore,} a change in its constraining power would be easier to appreciate. We should mention at this point that the RSD also impacts the XC part of the 3$\times$2pt {because} the latter contains all the combinations of GC, WL, and XC. However, we {did} not perform the constraining power test due to the RSD in the 3$\times$2pt {because} the WL contribution which is not affected by RSD would make the RSD impact less evident. We again {focused} on the $w_0w_a$CDM model and the pessimistic scale cut. This is reasonable {because} RSD mostly contributes on scales $\ell<100$ (see \Cref{fig:RSD_bins}), and therefore a higher $\ell_\text{max}$ cut would not affect the constraining power for this test.

In particular, we {compared} the following two models. For the first model, we {constructed} the synthetic data set and covariance matrix for the photometric GC spectra including RSD (GC with RSD), and we fit it against the predictions of the $w_0w_a$CDM model, including RSD. The results of this analysis are shown by the orange contours in the bottom panel of \Cref{fig:Fisher_MCMC_constpow}. For the second model, we {did} exactly the same, but neither the synthetic data along with the covariance matrix nor the theory model {included} the RSD correction (GC without RSD); this corresponds to the green contours in the bottom panel of \Cref{fig:Fisher_MCMC_constpow}.

It is clear that the constraints of {the two} models agree very well, indicating that the RSD correction in our modelling does not add significant information on the projected cosmological parameters. This means that {even though the} RSD account for up to $40\%$ of the signal for some redshift bins on the largest scales (see again top panel of \Cref{fig:RSD_bins}), the cosmological information does not come from the large scales {because they are dominated by cosmic variance, but gradually include smaller scales, where the RSD signal contribution becomes progressively less important}.

\begin{table*}
    \centering
    \caption{Summary of the mean estimated values, $\theta^\ast$, along their corresponding $68\%$ C.L.\ intervals, $\sigma_{\theta}$, and the relative bias, $B_\theta$ (see Eq.~\ref{eq:relative_bias}) of a given parameter.}
	\begin{tabularx}{\textwidth}{XXcccccccc}
           \hline
      & & \multicolumn{3}{c}{w/ RSD} && \multicolumn{3}{c}{w/o RSD}\\
       \cline{3-5}\cline{7-9}
      Parameter & Scale cut & \multicolumn{1}{c}{$\theta^\ast$} & \multicolumn{1}{c}{$\sigma_{\theta}$} & \multicolumn{1}{c}{$B_\theta$} && \multicolumn{1}{c}{$\theta^\ast$} & \multicolumn{1}{c}{$\sigma_{\theta}$} & 
      \multicolumn{1}{c}{$B_\theta$} \\
           \hline
       \multirow{2}{*}{$\om$} & pess. & $0.3193$ & $0.0052$ & 0.134 && \ \ $0.3019$ & $0.0044$ & \textbf{4.11} \\
   & opt. & $0.3197$ & $0.0037$ & 0.08 && \ \ $0.3022$ & $0.0033$ & \textbf{5.39} \\
    \hline
    \multirow{2}{*}{$\ob$} & pess. & $0.0503$ & $0.0024$ & 0.125 && \ \ $0.0499$ & $0.0022$ & 0.045 \\
   & opt. & $0.0502$ & $0.0018$ & 0.111 && \ \ $0.0506$ & $0.0018$ & 0.33 \\
    \hline
    \multirow{2}{*}{$h$} & pess. & $0.674$ & $0.023$ & 0.174 && \ \ $0.668$ & $0.021$ & 0.095 \\
   & opt. & $0.672$ & $0.015$ & 0.133 && \ \ $0.673$ & $0.016$ & 0.18 \\
    \hline
    \multirow{2}{*}{$\sigma_8$} & pess. & $0.8154$ & $0.0059$ & 0.101 && \ \ $0.8327$ & $0.0062$ & \textbf{2.69} \\
   & opt. & $0.816$ & $0.0013$ & 0.0 && \ \ $0.8206$ & $0.0013$ & \textbf{3.53} \\
    \hline
    \multirow{2}{*}{$n_{\rm s}$} & pess. & $0.96$ & $0.012$ & 0.0 && \ \ $0.959$ & $0.011$ & 0.091 \\
   & opt. & $0.9596$ & $0.0037$ & 0.108 && \ \ $0.9684$ & $0.0038$ & \textbf{2.21} \\
    \hline
    \multirow{2}{*}{$S_8$} & pess. & $0.8412$ & $0.0095$ & 0.164 && \ \ $0.8353$ & $0.009$ & 0.829 \\
   & opt. & $0.8424$ & $0.0046$ & 0.078 && \ \ $0.8236$ & $0.0043$ & \textbf{4.45} \\
    \hline
    \smallskip
	\end{tabularx}
    \label{table:results_GC_LCDM}
{\small
\raggedright \textit{Note:} These are the results of the GC analysis for each $\Lambda$CDM cosmological parameter for the complete (with RSD) and the incomplete (without RSD) model and for the pessimistic and optimistic scale cuts. We highlight the most biased cases ($B_\theta>1$) with \textbf{bold}. \par}
\end{table*}

\begin{table*}
    \centering
    \caption{Same as \Cref{table:results_GC_LCDM} for GC but for the $w_0w_a$CDM model.}
	\begin{tabularx}{\textwidth}{XXccccccccc}
           \hline
      & & \multicolumn{3}{c}{w/ RSD} && \multicolumn{3}{c}{w/o RSD}\\
       \cline{3-5}\cline{7-9}
      Parameter & Scale cut & \multicolumn{1}{c}{$\theta^\ast$} & \multicolumn{1}{c}{$\sigma_{\theta}$} & \multicolumn{1}{c}{$B_\theta$} && \multicolumn{1}{c}{$\theta^\ast$} & \multicolumn{1}{c}{$\sigma_{\theta}$} & 
      \multicolumn{1}{c}{$B_\theta$} \\
           \hline
      $\om$
   & opt. & $0.3190$ & $0.0069$ & 0.145 && \ \ $0.3092$ & $0.0056$ & \textbf{1.928} \\
    \hline
    $\ob$
   & opt. & $0.0499$ & $0.0019$ & 0.052 && \ \ $0.0516$ & $0.0019$ & 0.84 \\
    \hline
    $h$
   & opt. & $0.672$ & $0.018$ & 0.11 && \ \ $0.664$ & $0.016$ & 0.37 \\
    \hline
    $\sigma_8$
   & opt. & $0.8166$ & $0.003$ & 0.2 && \ \ $0.8158$ & $0.0027$ & 0.074 \\
    \hline
    $n_{\rm s}$
   & opt. & $0.96$ & $0.0043$ & 0.0 && \ \ $0.9669$ & $0.0043$ & \textbf{1.60} \\
    \hline
    $w_0$
   & opt. & $-1.001$ & $0.042$ & 0.0238 && \ \ $-1.007$ & $0.039$ & 0.17 \\
   \hline
    $w_a$
   & opt. & $-0.02$ & $0.18$ & 0.111 && \ \ $0.21$ & $0.17$ & \textbf{1.235} \\
   \hline
    $S_8$
   & opt. & $0.842$ & $0.007$ & 0.108 && \ \ $0.8281$ & $0.0058$ & \textbf{2.52} \\
   \hline
	\end{tabularx}
    \label{table:results_GC_w0waCDM}
\end{table*}

\subsection{Ignoring RSD}
\label{sec:biases}
Regardless of the absence of {an} additional constraining power on cosmological parameters encoded in linear RSD, we {proceeded to investigate the} effect of neglecting RSD in our modelling. {This} investigation was performed with an MCMC analysis in \citet{Tanidis:2019teo}, assuming a \Euclid-like survey of {intermediate-width} Gaussian bins (with a photometric redshift scatter $0.05(1+z)$). {We found} that it is crucial to include RSD in the theoretical predictions {to avoid biasing} the cosmological parameters. This outcome of the importance of RSD in the modelling {agreed} with the decision {in previous studies} to include them in {the} analyses (see again the references in \Cref{sec:intro}). Following the same approach, we {created a} mock data vector and covariance for the \Euclid survey, with the same specifications as in \citetalias{ISTF}, including the RSD contribution (\Cref{sec:Limber_RSDs}). The data {were} then analysed assuming an incorrect model {without} the RSD contribution. The impact of {this} approximation on the accuracy of the final constraints depends on the amplitude of the RSD signal, but also on the details of the experimental noise. {We therefore extended} the investigation of \citet{Tanidis:2019teo}{, which was only performed} for GC in a $\Lambda$CDM model and in the linear regime \footnote{In \citet{Tanidis:2019teo} only the linear matter power spectrum was used and $\ell_\text{max}$ was defined by $k_\text{max}\,\chi(\bar{z})$ with $k_\text{max}\simeq0.25\,h\,\mathrm{Mpc}^{-1}$ and $\bar{z}$ the mean redshift of the bin.} for the density field, to more realistic settings. That is, {we still considered} the linear RSD, but now in the non-linear regime of the density field. In addition, we {exploited} photometric redshift bins from the Flagship simulation, including massive neutrinos in the analysis, and {also investigated the effect} on the $w_0w_a$CDM model. Finally, we {performed} the analysis both for GC alone and for the full 3$\times$2pt, using the pessimistic and optimistic multipole cuts for both cases.

Finally, to {gain better insight into the degeneracies} that are present in the incorrect modelling, we recast all the final constraints on $\om$ and $\sigma_8$ in the derived parameter $\smash{S_8=\sigma_8\,\sqrt{\om/0.3}}$. This parameter is particularly informative about the degenerate direction between $\om$ and $\sigma_8$ in the {WL} measurements that are included in the 3$\times$2pt. However, for the sake of completeness, {we calculated} it for the GC {alone} as well.


\subsubsection{Biased parameter constraints in the photometric GC}
\label{sec:biased_GC}
We {started} by obtaining constraints for both the correct (den+RSD) and incorrect models (den-only), using photometric GC alone, both assuming $\Lambda$CDM and its extension $w_0w_a$CDM. To assess the bias on the parameters of interest {when} an incorrect model is assumed, we {used} the relative bias which is valid for deterministic quantities assuming no stochasticity (e.g. a noiseless data vector {such as} we consider here),
\begin{equation}
B_\theta=\frac{|\theta^\ast-\theta^{\text{fid}}|}{\sigma_\theta}\;,
    \label{eq:relative_bias}
\end{equation}
with $\theta^\ast$ being the mean estimated value on the marginalised posterior of the parameter $\theta$, $\sigma_\theta$ the corresponding $68\%$ C.L., and $\theta^{\rm \text{fid}}$ the input fiducial value. It is defined as the offset we {obtain} from the input fiducial value (our benchmark) given the expected uncertainty on the parameter. \citet{Massey2012} {suggested that after taking systematic effects into account, the} values of $B_\theta\gtrsim0.3$ can be considered already as statistically significant \footnote{Values of $B_\theta<0.3$ are considered to be permissible within the statistical fluctuation assuming that the same information for the synthetic data and the theory modelling {was} used.}.

\begin{figure*}
    \centering
    \includegraphics[width=0.65\textwidth]{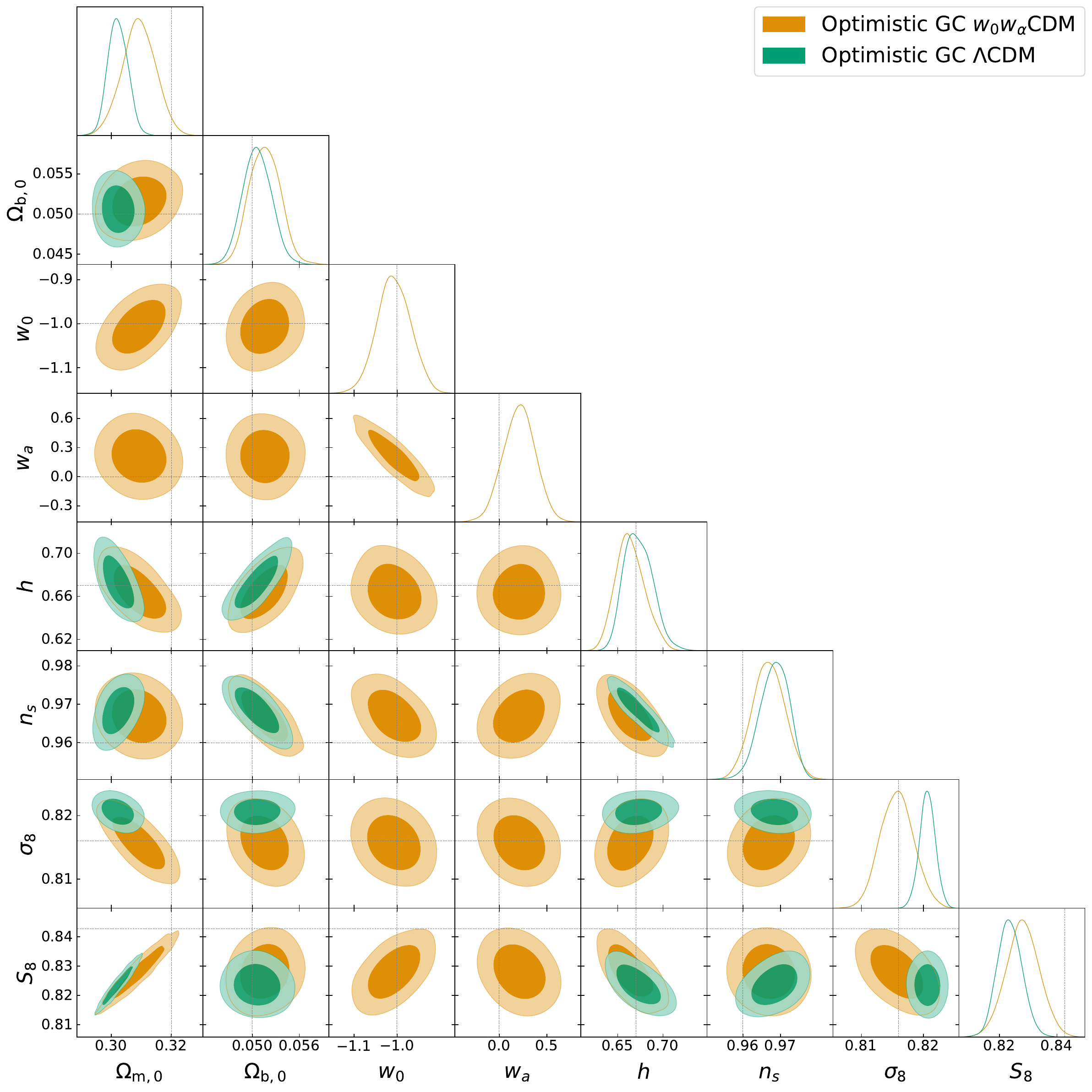}\\
    \includegraphics[width=0.65\textwidth]{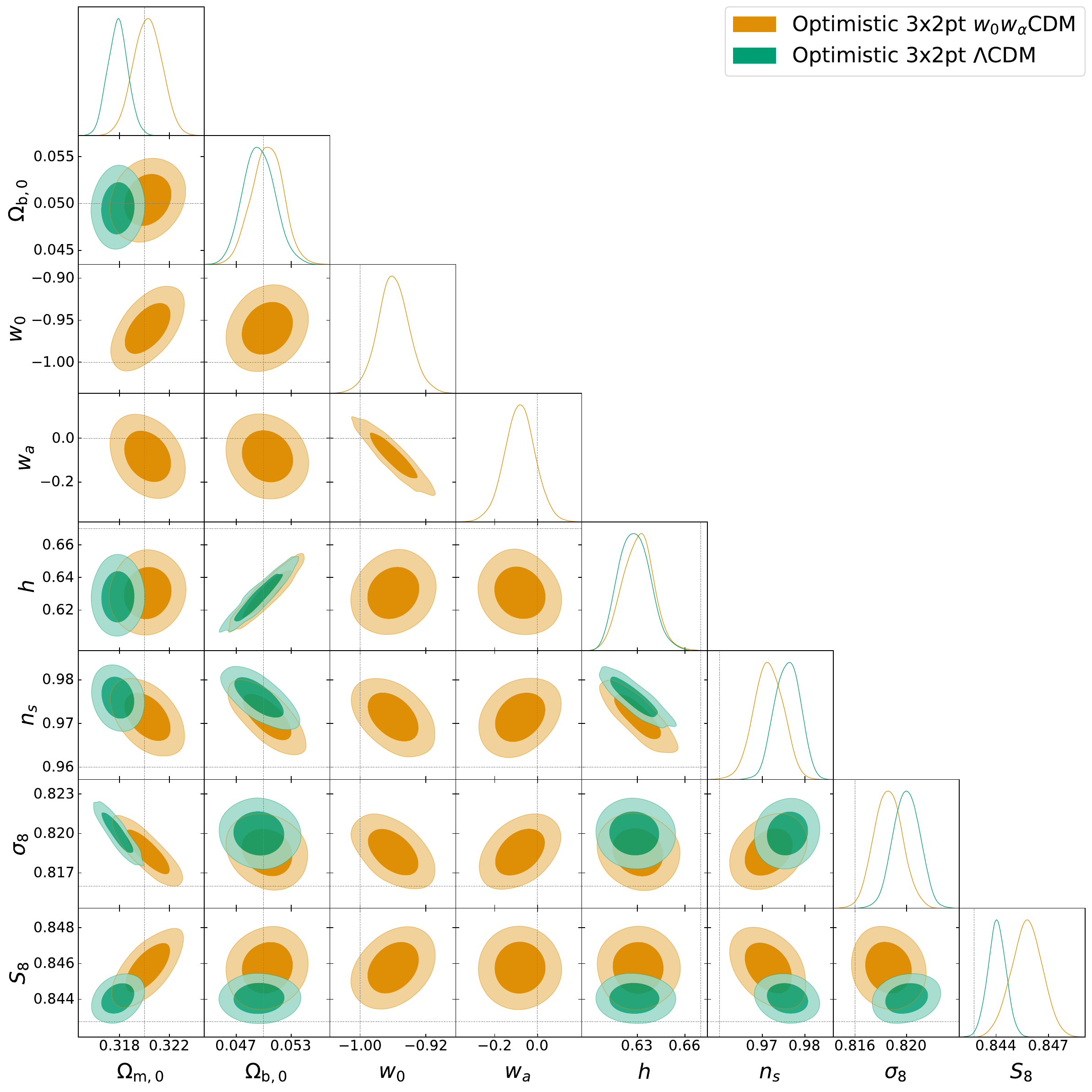}
    \caption{$68\%$ and $95\%$ C.L.\ marginalised contours alongside the corresponding one-dimensional posterior distributions of the cosmological parameters for the incorrect modelling (without RSD). \textit{Top:} Constraints from the GC for the $\Lambda$CDM (green contours) and $w_0w_a$CDM model (orange contours) {for} the optimistic scale cut. The fiducial cosmology is marked by dotted black lines. \textit{Bottom:} Same as above, but for 3$\times$2pt.}
    \label{fig:GC_3x2pt_opt_biased}
\end{figure*}

All the $\Lambda$CDM results in GC for the correct and incorrect model, as well as the pessimistic and optimistic scale cuts, are shown in \Cref{table:results_GC_LCDM}. {We opted to present the} incorrect model constraints for the optimistic cases alone {in the top panel of \Cref{fig:GC_3x2pt_opt_biased} because the} resulting biases are larger than those in the pessimistic cut.
We {note that} in the pessimistic case, neglecting RSD leads to a bias of $4.11$ and $2.69$ on {the} parameters $\om$ and $\sigma_8$, respectively; while with the correct modelling we always recover the fiducial cosmology with $B_\theta\lesssim0.3$ (see again \Cref{table:results_GC_LCDM}). This bias is also imprinted on the increase {in} the best-fit value by $\Delta\chi^2\approx84.61$ with respect to the correct model.

The picture is similar for the optimistic scale cut, as the green contours {in} the top panel of \Cref{fig:GC_3x2pt_opt_biased} {show. We} note that the bias on the same parameters increases, and another mild {bias} now  arises on $n_{\rm s}$. In particular, {the} parameters $\om$, $\sigma_8$, and $n_{\rm s}$ are biased with values $5.39$, $3.53$, and $2.21$, respectively, and as expected there is increase {in} $\Delta\chi^2\approx92.48$ with respect to the correct model. This bias increase in the optimistic compared to the pessimistic {scenario} can be explained as follows: {By} extending the range to higher multipoles, we do not include more signal from the linear RSD which, as we saw in \Cref{sec:Limber_RSDs}, contributes only on large scales ($\ell\lesssim100$). On the other hand, we increase the constraining power on the parameters, meaning that any existing {bias} is enhanced by lowering the projected errors. {This} shows that an inaccurate modelling becomes progressively more crucial {with improving data quality.}

We {now turn our attention on} the results obtained assuming the $w_0w_a$CDM model. Again, we only {considered} the optimistic scales cut. The reason for this is that GC alone for the pessimistic scale cut is not very constraining in this model, and the bias on the parameters is expected to be small due to {the smaller range of scales and the larger parameter set.} The result of all this is that the $w_0$ and $w_a$ parameters vary over a {wide} range in the MCMC, also in regions where $w_0+w_a$ can take non-negative values \citep[see][ for details on the high-redshift limit of $w(z)$ in CPL]{Cepa}. In {this} region of the parameter space, \camb\ cannot obtain meaningful cosmological quantities, and these points are automatically rejected from the chain, thus introducing a cut in the parameter space {that} is physically motivated. Based on this restriction {and} {because the} modelling for this case is not very constraining{, regardless} of the assumed prior on the CPL parameters, we decided to focus our analysis {on} the optimistic case {alone}. The results are shown in \Cref{table:results_GC_w0waCDM} and {in} the top panel of \Cref{fig:GC_3x2pt_opt_biased}, where we still have biased estimates of our free parameters, {but with a lower significance than in} the $\Lambda$CDM analysis {because the} uncertainty brought by keeping the dark energy parameters $w_0$ and $w_a$ free {is larger}. {The parameters that are} shifted by more than $1\sigma$ are $\om$, $n_{\rm s}$, and $w_a$, with their $B_\theta$ values being $1.9$, $1.6$, and $1.2$, respectively. Again, there is an increase {in} $\Delta\chi^2\approx90.65$ compared to the complete model. We summarise all the $B_\theta$ values for the aforementioned scenarios in the left panel of \Cref{fig:summary_plot} (see {the} caption for details), where {the trend is clearer} for larger biases in the optimistic compared to the pessimistic {cases and for} those of the $\Lambda$CDM model against the $w_0w_a$CDM extension.
\begin{table*}
    \centering
    \caption{Same as \Cref{table:results_GC_LCDM} but for the 3$\times$2pt.}
	\begin{tabularx}{0.9\textwidth}{XXccccccccc}
           \hline
      & & \multicolumn{3}{c}{w/ RSD} && \multicolumn{3}{c}{w/o RSD}\\
       \cline{3-5}\cline{7-9}
      Parameter & scale cut & \multicolumn{1}{c}{$\theta^\ast$} & \multicolumn{1}{c}{$\sigma_{\theta}$} & \multicolumn{1}{c}{$B_\theta$} && \multicolumn{1}{c}{$\theta^\ast$} & \multicolumn{1}{c}{$\sigma_{\theta}$} & 
      \multicolumn{1}{c}{$B_\theta$} \\
           \hline
       \multirow{2}{*}{$\om$} & pess. & $0.32$ & $0.0025$ & 0.0 && \ \ $0.3119$ & $0.0023$ & \textbf{3.52} \\
   & opt. & $0.32006$ & $0.00082$ & 0.07317 && \ \ $0.31786$ & $0.00082$ & \textbf{2.61} \\
    \hline
    \multirow{2}{*}{$\ob$} & pess. & $0.05$ & $0.0022$ & 0.0 && \ \ $0.0481$ & $0.002$ & 0.95 \\
   & opt. & $0.0501$ & $0.0018$ & 0.0555  && \ \ $0.0495$ & $0.0017$ & 0.29 \\
    \hline
    \multirow{2}{*}{$h$} & pess. & $0.671$ & $0.016$ & 0.062 && \ \ $0.625$ & $0.014$ & \textbf{3.21} \\
   & opt. & $0.671$ & $0.011$ & 0.090 && \ \ $0.6283$ & $0.01015$ & \textbf{4.11} \\
    \hline
    \multirow{2}{*}{$\sigma_8$} & pess. & $0.816$ & $0.0032$ & 0.0 && \ \ $0.8279$ & $0.003$ & \textbf{3.96} \\
   & opt. & $0.816$ & $0.001$ & 0.0 && \ \ $0.82$ & $0.001$ & \textbf{4.0} \\
    \hline
    \multirow{2}{*}{$n_{\rm s}$} & pess. & $0.9596$ & $0.0078$ & 0.0512 && \ \ $0.9747$ & $0.0072$ & \textbf{2.05} \\
   & opt. & $0.9598$ & $0.0031$ & 0.0645 && \ \ $0.9759$ & $0.003$ & \textbf{5.3} \\
    \hline
    \multirow{2}{*}{$S_8$} & pess. & $0.8428$ & $0.00073$ & 0.0531 && \ \ $0.84422$ & $0.00072$ & \textbf{2.026} \\
   & opt. & $0.8428$ & $0.00054$ & 0.07189 && \ \ $0.84405$ & $0.00052$ & \textbf{2.478} \\
    \hline
	\end{tabularx}
    \label{table:results_3x2pt_LCDM}
\end{table*}

\begin{table*}
    \centering
    \caption{Same as \Cref{table:results_GC_w0waCDM} but for the 3$\times$2pt.}
	\begin{tabularx}{0.9\textwidth}{XXccccccccc}
           \hline
      & & \multicolumn{3}{c}{w/ RSD} && \multicolumn{3}{c}{w/o RSD}\\
       \cline{3-5}\cline{7-9}
      Parameter & scale cut & \multicolumn{1}{c}{$\theta^\ast$} & \multicolumn{1}{c}{$\sigma_{\theta}$} & \multicolumn{1}{c}{$B_\theta$} && \multicolumn{1}{c}{$\theta^\ast$} & \multicolumn{1}{c}{$\sigma_{\theta}$} & 
      \multicolumn{1}{c}{$B_\theta$} \\
           \hline
       \multirow{2}{*}{$\om$} & pess. & $0.3202$ & $0.0038$ & 0.0526 && \ \ $0.3159$ & $0.0035$ & \textbf{1.17} \\
   & opt. & $0.32$ & $0.0012$ & 0.0  && \ \ $0.3202$ & $0.0012$ & 0.16 \\
    \hline
    \multirow{2}{*}{$\ob$} & pess. & $0.0502$ & $0.0025$ & 0.0799 && \ \ $0.0520$ & $0.0024$ & 0.83 \\
   & opt. & $0.0499$ & $0.0019$ & 0.0526 && \ \ $0.0504$ & $0.0017$ & 0.23 \\
    \hline
    \multirow{2}{*}{$h$} & pess. & $0.671$ & $0.019$ & 0.0526 && \ \ $0.653$ & $0.017$ & 1.0 \\
   & opt. & $0.67$ & $0.01$ & 0.0 && \ \ $0.6306$ & $0.0098$ & \textbf{4.02} \\
    \hline
    \multirow{2}{*}{$\sigma_8$} & pess. & $0.8158$ & $0.0047$ & 0.0425 && \ \ $0.8272$ & $0.0044$ & \textbf{2.54} \\
   & opt. & $0.816$ & $0.0012$ & 0.0 && \ \ $0.8186$ & $0.0011$ & \textbf{2.36} \\
    \hline
    \multirow{2}{*}{$n_{\rm s}$} & pess. & $0.9595$ & $0.0094$ & 0.0531 && \ \ $0.9557$ & $0.0087$ & 0.49  \\
   & opt. & $0.9602$ & $0.0035$ & 0.0571 && \ \ $0.9715$ & $0.0035$ & \textbf{3.28} \\
    \hline
    \multirow{2}{*}{$w_0$} & pess. & $-0.9996$ & $0.046$ & 0.0086 && \ \ $-0.945$ & $0.044$ & \textbf{1.25} \\
   & opt. & $-0.999$ & $0.02$ & 0.05 && \ \ $-0.96$ & $0.02$ & \textbf{2.0} \\
   \hline
    \multirow{2}{*}{$w_a$} & pess. & $0.0$ & $0.18$ & 0.0 && \ \ $0.02$ & $0.16$ & 0.125 \\
   & opt. & $-0.005$ & $0.075$ & 0.066 && \ \ $-0.083$ & $0.073$ & \textbf{1.136} \\
   \hline
    \multirow{2}{*}{$S_8$} & pess. & $0.8428$ & $0.0014$ & 0.0277 && \ \ $0.8488$ & $0.0014$ & \textbf{4.31} \\
   & opt. & $0.84273$ & $0.00086$ & 0.0362 && \ \ $0.84576$ & $0.00089$ & \textbf{3.37} \\
   \hline
	\end{tabularx}
    \label{table:results_3x2pt_w0waCDM}
\end{table*}

\subsubsection{Biased parameter constraints in the 3$\times$2pt}
We {repeated} exactly the same analysis (different cosmologies and scale cuts) but now for the 3$\times$2pt (\Cref{table:results_3x2pt_LCDM}, \Cref{table:results_3x2pt_w0waCDM}, and {the} bottom panel of  \Cref{fig:GC_3x2pt_opt_biased}). Similarly to what we saw in \Cref{sec:biased_GC} for the $\Lambda$CDM model and the pessimistic scale cuts, the parameters $\om$, $h$, $n_{\rm s}$, and $\sigma_8$ are now biased by $3.52$, $3.21$, $2.05$, and $3.96$ (see \Cref{table:results_3x2pt_LCDM}), while the overall $\Delta\chi^2$ compared to the correct model is increased by 96.27. It is interesting to {note that} while some of the parameters ({e.g.} $n_{\rm s}$ and $h$) exhibit an increased $B_\theta$ value due to the increased constraining power of the 3$\times$2pt combination, for the other parameters the behaviour is less straightforward. On the one hand, WL strongly constrains $S_8${, which is} a combination of $\om$ and $\sigma_8${, and because} this probe is not biased by neglecting RSD, {we would expect a lower bias for these parameters}. On the other hand, however, the inclusion of XC lifts the degeneracy between $\sigma_8$ and the bias parameters present in the GC probe, thus making the latter more sensitive to this parameter. {Because} both GC and XC are biased when we neglect RSD, the overall effect is an increased $B_\theta$ value on $\sigma_8$ with respect to the {case of GC alone}.

In the optimistic case, the balance between these effects changes {because the increased number} of scales available for WL makes this more relevant, thus reducing the bias with respect to the {case of GC alone}. This does not apply to $n_{\rm s}$ and $h$, which are still mostly constrained by GC. The $\Delta\chi^2$ is increased by 106.78 with respect to the correct model.

\begin{figure*}
    \centering
    \includegraphics[width=1.0\textwidth]{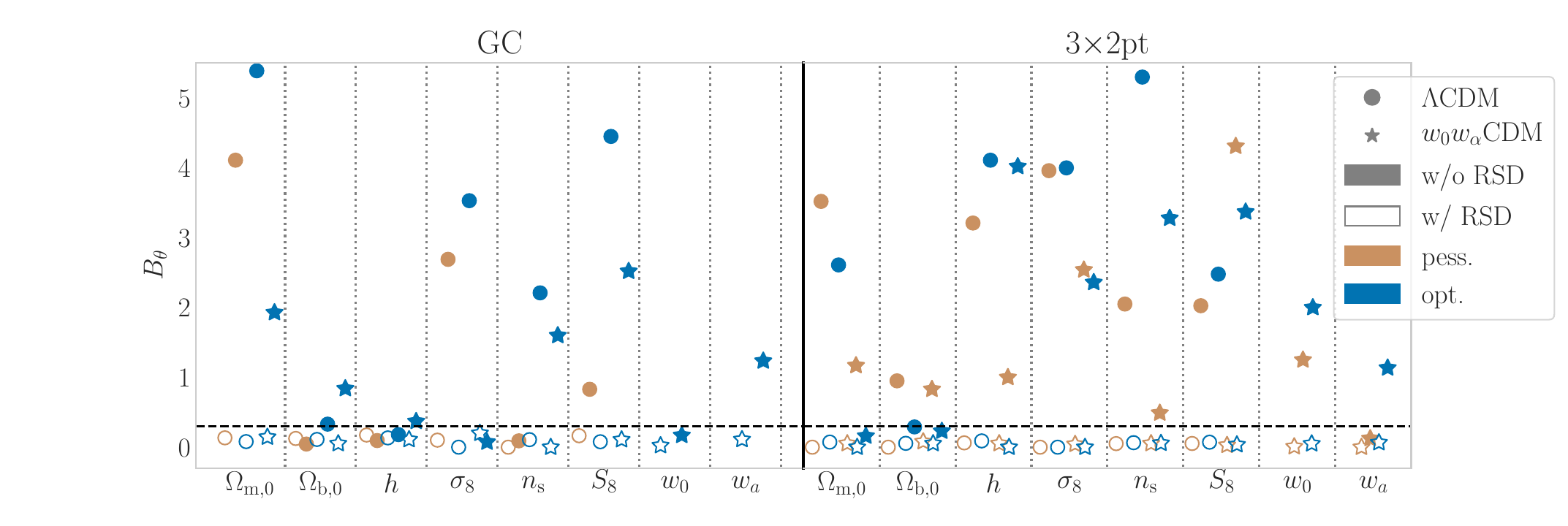}\\
    \caption{Summary plot of the $B_\theta$ values for the GC (left) and the 3$\times$2pt (right) for all the examined cases. The circles correspond to the $\Lambda$CDM cosmology model and the stars {show} the $w_0w_\alpha$CDM extension{. The} open and filled symbols {show} a theory modelling with and without RSD, respectively. {The pessimistic} and optimistic scale cuts are shown {in} brown and blue. The horizontal dashed line denotes the value $B_\theta=0.3$}
    \label{fig:summary_plot}
\end{figure*}

Finally, the results of the 3$\times$2pt and the $w_0w_a$CDM  model are shown in \Cref{table:results_3x2pt_w0waCDM}. In the pessimistic case, {the biased estimate for $\sigma_8$ is} $2.54$, and the peaks for the parameters $\om$, $\ob$, $w_0$, and $h$ {are misplaced with biases of} $1.17$, $0.83$, $1.25$, and $1.00$, while for the optimistic scenario (see again \Cref{table:results_3x2pt_w0waCDM} and the orange contours {in} the bottom panel of \Cref{fig:GC_3x2pt_opt_biased}), there are biased estimates for $w_0$, $w_a$, $h$, $n_{\rm s}$, and $\sigma_8$ at $2.00$, $1.14$, $4.02$, $3.28$, and $2.36$, respectively. Both models yield an increased $\Delta\chi^2$ compared to the complete model by 92.05 (pessimistic) and 103.6 (optimistic). When the latter case {is compared} to GC alone, the increased constraining power, brought by WL and XC, has the effect of increasing the significance of the bias on most of the parameters. Similarly to \Cref{sec:biased_GC}, we summarise the $B_\theta$ values for all the cases of the 3$\times$2pt in the right panel of \Cref{fig:summary_plot}.

\section{Conclusions}\label{sec:conclusions}
In this work, we have aimed to quantify the contribution of linear {RSD in photometric GC as is} expected to be measured by \Euclid, both as a stand-alone probe and in combination with cosmic shear (WL), in the so-called 3$\times$2pt approach. We followed \citet{Tanidis:2019teo} {and included}  RSD in the angular power spectra of GC and its cross-correlation with WL (XC).

Using the galaxy distribution information coming from the Flagship simulation, and the \Euclid specifications discussed in \citetalias{ISTF}, we generated synthetic data by generating angular power spectra and covariance matrix with a fiducial cosmology, for the photometric observations of \Euclid and produced the posterior distributions for the free parameters of our model in {an MCMC} framework.

As a first step, we validated our results against a Fisher matrix approach {and found that they agree very well}. Then, we compared the constraints on {the cosmological parameters that are} obtained when the theoretical predictions are computed {with} the contribution of RSD to those obtained when RSD are neglected. {When GC is used} as a stand-alone probe {and} a $\Lambda$CDM cosmology {is assumed} for the theoretical predictions, neglecting RSD can lead to significant inaccuracies on the reconstruction of cosmological parameters{, in particular,} in the most optimistic case {when} the constraining power of our experimental setup is maximum. We find that {the} parameters $\om$, $n_{\rm s}$, and $\sigma_8$ are all significantly shifted from their (input) fiducial values with biases of $5.4\,\sigma$, $2.2\,\sigma$, and $3.5\,\sigma${, respectively}. The statistical significance of these shifts is reduced when the cosmological model used to fit the data allows {the} $w_0$ and $w_a$ parameters to take values different from $\Lambda$CDM{, however}. The inclusion of these two {additional} parameters degrades the constraining power and leads to a less evident shift {than in} the fiducial model.

{We} included WL and XC in the analysis {to perform} a 3$\times$2pt analysis. In this case, we found a non-trivial effect on the significance of the RSD contribution. On the one hand, WL contributes to tighten the constraints {that can be achieved} with \Euclid, thus potentially increasing the significance of the shifts {that are obtained when RSD is neglected}. On the other hand, {the} theoretical predictions for this probe are not biased by {this} approximation (although the XC is still biased), and the inclusion of WL can therefore drag the recovered posterior distribution towards the fiducial values even when RSD are neglected. {We} found that for parameters {such as} $n_{\rm s}$ and $h$, where galaxy GC dominate, WL simply improve the constraining power, and the bias on these parameters increases in significance. {In contrast, the parameters that are} mostly constrained by WL, such as $\om$ and $\sigma_8$, we found that their shifts decrease in the optimistic case, where the {constraining power} of WL dominates the constraining power from GC.

{To summarise}, we found that {when} the contribution of linear RSD {is not included} in the theoretical predictions for GC angular power spectra, {it can significantly reduce the accuracy but not the precision of the constraints that can be achieved by} \Euclid. {The reason is} that the linear RSD contribute at scales $\ell<100$, {which is} a cosmic-variance dominated regime {that yields} no gain in cosmological information. It is important to note that the necessity {of including} the effect of linear RSD in GC in order to avoid cosmology biases has been studied in depth \citep{Fisher1993, Heaven1995, Padmanabhan, Blake, Nock,Crocce, Balaguera-Antolinez2018, Abbott21} and the findings of our work agree with this. However, we have demonstrated that the approach of \citet{Tanidis:2019teo}, which is a fast and approximated way to account for the linear RSD correction, can {easily be} implemented and tested within a parameter estimation pipeline in the modelling of \Euclid photometric observables. In addition to this, {work is ongoing} to improve the modelling and study the effect of the non-linear galaxy bias in the photometric GC, which becomes especially important for Stage-IV galaxy surveys {such as} \Euclid.

\begin{acknowledgements}
KT is supported by the STFC grant ST/W000903/1 and by the European Structural and Investment Fund. For most of the development of this project KT was supported by the Czech Ministry of Education, Youth and Sports (Project CoGraDS - CZ.02.1.01/0.0/0.0/15\_003/0000437). V.C and M.M. acknowledge funding by the Agenzia Spaziale Italiana (\textsc{asi}) under agreement no. 2018-23-HH.0 and support from INFN/Euclid Sezione di Roma. IT acknowledges funding from the European Research Council (ERC) under the European Union's Horizon 2020 research and innovation programme (Grant agreement No. 863929; project title ``Testing the law of gravity with novel large-scale structure observables''). SC acknowledges support from the `Departments of Excellence 2018-2022' Grant (L.\ 232/2016) awarded by the Italian Ministry of University and Research (\textsc{mur}).
\AckEC
\end{acknowledgements}


\bibliographystyle{aa}
\bibliography{Biblio}

\appendix


\end{document}